%% file: buddy.tex
\documentclass{mypaper}

\input{font}

\usepackage{enumitem}

\usepackage{fancyhdr}
\fancypagestyle{firstpage}{
  \fancyhf{}
  \fancyhead[C]{\normalsize{HPCA 2017 Submission \#119 -- Confidential Draft -- Do NOT distribute!!}} 
}

\title{\sffamily Buddy-RAM: Improving the Performance and
  Efficiency of\\ Bulk Bitwise Operations Using DRAM}

\author{Vivek Seshadri$^1$, Donghyuk Lee$^2$, Thomas Mullins$^3$, Hasan
  Hassan$^4$, Amirali Boroumand$^5$,\\Jeremie Kim$^5$, Michael A. Kozuch$^6$, Onur
  Mutlu$^7$, Phillip B. Gibbons$^5$, Todd C. Mowry$^5$\\
  \small{\sffamily $^1$Microsoft Research $^2$NVIDIA Research $^3$Intel
  $^4$TOBB University of Economics and
    Technology}\\\small{\sffamily $^5$Carnegie
  Mellon University $^6$Intel Pittsburgh $^7$ETH Zurich}}
\date{}

\input{tikz-stuff}
\input{acronyms}

\begin{document}
\maketitle

\titlespacing{\subsubsection}{0pt}{5pt}{2pt}
\titlespacing{\subsection}{0pt}{5pt}{2pt}
\titlespacing{\section}{0pt}{9pt}{5pt}

\addtolength{\intextsep}{-2mm}
\addtolength{\textfloatsep}{-5mm}
\addtolength{\dbltextfloatsep}{-3mm}

\thispagestyle{empty}
\begin{abstract}
  \normalfont
  \input{abstract}

\end{abstract}

\setstretch{0.95}
\input{introduction}

\input{background}
\input{mechanism}
\input{implementation}

\input{support}
\input{lte-analysis}
\input{applications}

\input{related}
\input{conclusion}

\begin{footnotesize}
\bibliographystyle{plain}
\bibliography{references}
\end{footnotesize}

\end{document}

%% file: font.tex
\usepackage[T1]{fontenc} 
\usepackage{times}

%% file: tikz-stuff.tex
\usepackage{tikz}
\usetikzlibrary{arrows}
\usetikzlibrary{patterns}
\usetikzlibrary{calc}
\usetikzlibrary{shapes}
\usetikzlibrary{positioning,fit}

\usepackage{circuitikz}

%% file: acronyms.tex
\newcommand{\vdd}{V$_{DD}$\xspace}
\newcommand{\halfvdd}{$\frac{1}{2}$V$_{DD}$\xspace}

\newcommand{\halfvddpd}{$\frac{1}{2}$V$_{DD}+\delta$\xspace}

\newcommand{\cmdact}{\texttt{ACTIVATE}\xspace}
\newcommand{\cmdpre}{\texttt{PRECHARGE}\xspace}
\newcommand{\cmdwr}{\texttt{WRITE}\xspace}
\newcommand{\cmdrd}{\texttt{READ}\xspace}

\newcommand{\cmdacts}{\texttt{ACTIVATE}s\xspace}

\newcommand{\bbar}{$\overline{\textrm{bitline}}$\xspace}

\newcommand{\band}{\textrm{\texttt{and}}\xspace}
\newcommand{\bor}{\textrm{\texttt{or}}\xspace}
\newcommand{\bnot}{\textrm{\texttt{not}}\xspace}
\newcommand{\bxor}{\textrm{\texttt{xor}}\xspace}
\newcommand{\bnand}{\textrm{\texttt{nand}}\xspace}
\newcommand{\bnor}{\textrm{\texttt{nor}}\xspace}
\newcommand{\bxnor}{\textrm{\texttt{xnor}}\xspace}

\newcommand{\dwordline}{\emph{d-wordline}\xspace}
\newcommand{\nwordline}{\emph{n-wordline}\xspace}

\newcommand{\baddr}[1]{\texttt{B#1}\xspace}
\newcommand{\taddr}[1]{\texttt{T#1}\xspace}
\newcommand{\daddr}[1]{\texttt{D#1}\xspace}

\newcommand{\dccdz}{\texttt{DCC0}\xspace}
\newcommand{\dccnz}{$\overline{\textrm{\texttt{DCC0}}}$\xspace}

\newcommand{\dccdo}{\texttt{DCC1}\xspace}
\newcommand{\dccno}{$\overline{\textrm{\texttt{DCC1}}}$\xspace}

\newcommand{\czero}{\texttt{C0}\xspace}
\newcommand{\cone}{\texttt{C1}\xspace}

\newcommand{\aap}{\texttt{AAP}\xspace}

\newcommand{\ap}{\texttt{AP}\xspace}

\newcommand{\tras}{t$_{\textrm{\footnotesize{RAS}}}$\xspace}
\newcommand{\trp}{t$_{\textrm{\footnotesize{RP}}}$\xspace}

\newcommand{\buddyao}{Buddy AND/OR\xspace}
\newcommand{\buddynot}{Buddy NOT\xspace}

\newcommand{\tra}{TRA\xspace}

%% file: abstract.tex
Bitwise operations are an important component of modern day
programming. Many widely-used data structures (e.g., bitmap
indices in databases) rely on fast bitwise operations on large bit
vectors to achieve high performance. Unfortunately, in existing
systems, regardless of the underlying architecture (e.g., CPU,
GPU, FPGA), the throughput of such bulk bitwise operations is
limited by the available memory bandwidth.

We propose Buddy, a new mechanism that exploits the analog
operation of DRAM to perform bulk bitwise operations completely
inside the DRAM chip, thereby not wasting any memory
bandwidth. Buddy consists of two components. First, simultaneous
activation of three DRAM rows that are connected to the same set
of sense amplifiers enables us to perform bitwise AND and OR
operations. Second, the inverters present in each sense amplifier
enables us to perform bitwise NOT operations, with modest changes
to the DRAM array. These two components, along with RowClone, a
prior proposal for fast row copying inside DRAM, make Buddy
functionally complete, thereby allowing it to perform any bitwise
operation efficiently inside DRAM. Our implementation of Buddy
largely exploits the existing DRAM structure and interface, and
incurs low overhead (1\% of DRAM chip area).

Our evaluations based on SPICE simulations show that, across seven
commonly-used bitwise operations, Buddy provides between
10.9X---25.6X improvement in raw throughput and 25.1X---59.5X
reduction in energy consumption.  We evaluate three real-world
data-intensive applications that exploit bitwise
operations. First, Buddy improves performance of database queries
that use bitmap indices for fast analytics by 6.0X compared to a
state-of-the-art baseline using SIMD operations. Second, Buddy
accelerates BitWeaving, a recently-proposed technique for fast
database scans, by 7.0X on average across a wide range of scan
parameters.  Third, for the commonly-used \emph{set} data
structure, Buddy improves performance of set intersection, union,
and difference operations by 3.0X compared to conventional
implementations. We also describe four other promising
applications that can benefit from Buddy, including DNA sequence
analysis, encryption, and approximate statistics.  We believe and
hope that the large performance and energy improvements provided
by Buddy can enable many other applications to use bitwise
operations.

%% file: introduction.tex
\section{Introduction}
\label{sec:introduction}

Bitwise operations are an important component of modern day
programming~\cite{btt-knuth,hacker-delight}. In this paper, we aim
to improve the performance and efficiency of bitwise operations on
large amounts of data, or \emph{bulk bitwise operations}, which
are triggered by many applications. For instance, in databases,
bitmap indices~\cite{bmide,bmidc}, which heavily use bitwise
operations, were shown to be more efficient than B-trees for many
queries~\cite{bmide,fastbit,bicompression}. Many real-world
databases~\cite{oracle,redis,fastbit,rlite} support bitmap
indices. Again in databases, a recent work,
BitWeaving~\cite{bitweaving}, proposes a technique to accelerate
database {\em scans} completely using bitwise operations. In
bioinformatics, prior works have proposed techniques to exploit
bitwise operations to accelerate DNA sequence
alignment~\cite{bitwise-alignment,shd}. Bitwise operations are
also prevalent in encryption algorithms~\cite{xor1,xor2,enc1},
graph processing~\cite{pinatubo}, approximate
statistics~\cite{summingbird}, and networking
workloads~\cite{hacker-delight}. Thus, accelerating such bulk
bitwise operations can significantly boost the performance and
efficiency of a number of important applications.

In existing systems, a bulk bitwise operation requires a large
amount of data to be transferred back and forth on the memory
channel between main memory and the processor. Such large data
transfers result in high latency, bandwidth, and energy
consumption. In fact, our experiments on an Intel
Skylake~\cite{intel-skylake} system and an NVIDIA GeForce GTX
745~\cite{gtx745} system show that the available memory bandwidth
of these systems is the performance bottleneck that limits the
throughput of bulk bitwise operations
(Section~\ref{sec:lte-analysis}).

In this paper, we propose a new mechanism to perform bulk bitwise
operations \emph{completely} inside main memory (DRAM), without
wasting memory bandwidth, improving both performance and
efficiency. We call our mechanism \emph{BUdDy-RAM}
(\textbf{B}itwise operations \textbf{U}sing \textbf{Dy}namic
\textbf{RAM}) or just \emph{Buddy}.  Buddy consists of two parts,
\emph{Buddy-AND/OR} and \emph{Buddy-NOT}. Both parts rely on the
operation of the \emph{sense amplifier} that is used to extract
data from the DRAM cells.

{\bf Buddy-AND/OR} exploits the fact that a DRAM chip consists of
many subarrays~\cite{salp,dsarp,rowclone}. In each subarray, many
rows of DRAM cells (typically 512 or 1024) share a single row of
sense amplifiers. We show that simultaneously activating three
rows (rather than one) results in a bitwise majority function,
i.e., in each column of cells, at least two cells have to be fully
charged for the corresponding sense amplifier to detect a logical
``1''. We refer to this operation as \emph{triple-row
  activation}. We show that by controlling the initial value of
one of the three rows, we can use the triple-row activation to
perform a bitwise AND or OR operation of the remaining two rows.
{\bf Buddy-NOT} exploits the fact that each sense amplifier
consists of two inverters. We use a \emph{dual-contact cell} (a
2-transistor 1-capacitor cell~\cite{2t-1c-1}) that connects to
both sides of the inverters to efficiently perform a bitwise NOT
of the value of any cell connected to the sense amplifier. Our
SPICE simulations results show that both Buddy-AND/OR and
Buddy-NOT work reliably, even in the presence of significant
process variation.  Sections~\ref{sec:bitwise-and-or} and
\ref{sec:bitwise-not} present these two components in full detail.

Combining Buddy-AND/OR and Buddy-NOT, Buddy is functionally
complete, and can perform \emph{any} bitwise logical
operation. Since DRAM internally operates at row granularity, both
Buddy-AND/OR and Buddy-NOT naturally operate at the same
granularity, i.e., an entire row of DRAM cells (multiple kilobytes
across a module). As a result, Buddy can efficiently perform any
multi-kilobyte-wide bitwise operation.

A naive implementation of Buddy would lead to high cost and
complexity. For instance, supporting triple-row activation on any
three {\em arbitrary} DRAM rows requires the replication of the
address bus and the row decoder, as these structures have to
communicate and decode three addresses simultaneously. In this
work, we present a practical, low-cost implementation of Buddy,
which heavily exploits the existing DRAM operation and command
interface. First, our implementation allows the memory controller
to perform triple-row activation \emph{only} on a \emph{designated
  set} of three rows (chosen at design time) in each DRAM
subarray. To perform Buddy AND/OR operation on two arbitrary DRAM
rows, the data in those rows are first copied to the designated
set of rows that are capable of triple-row activation, and the
result of the operation is copied out of the designated rows to
the destination. We exploit a recent work,
RowClone~\cite{rowclone}, which enables very fast data copying
between two DRAM rows, to perform the required copy operations
efficiently. Second, we logically split the row decoder in each
DRAM subarray into two parts, one part to handle activations
related to only the \emph{designated} rows, and another to handle
activations of the regular data rows. Finally, we introduce a
simple address mapping mechanism that avoids any changes to the
DRAM command interface. In fact, our implementation introduces
\emph{no} new DRAM commands. Putting together all these
techniques, we evaluate the area cost of our proposal to be
equivalent to 10 DRAM rows per subarray, which amounts to less
than 1\% of DRAM area (as shown in
Section~\ref{sec:hardware-cost}).

By performing bulk bitwise operations completely in DRAM, Buddy
significantly improves the performance and reduces the energy
consumption of these operations. In fact, as Buddy does not
require any data to be read out of the DRAM banks, each individual
Buddy operation is contained entirely inside a DRAM bank. Since
DRAM chips have multiple DRAM banks to overlap the latencies of
different memory requests, we can perform multiple Buddy
operations concurrently {\em in different banks}. As a result, the
performance of Buddy scales linearly with the number of DRAM banks
in the memory system, and is not limited by the off-chip pin
bandwidth of the processor.  We evaluate these benefits by
comparing the raw throughput and energy of performing bulk bitwise
operations using Buddy to an Intel Skylake~\cite{intel-skylake}
system and an NVIDIA GTX 745~\cite{gtx745} system. Our evaluations
show that the bitwise operation throughput of both of these
systems is limited by the off-chip memory bandwidth. Averaged
across seven commonly-used bitwise operations, Buddy, even when
using only \emph{a single DRAM bank}, improves bitwise operation
throughput by 3.8X---9.1X compared to the Skylake, and 2.7X---6.4X
compared to the GTX 745. Buddy reduces DRAM energy consumption of
these bitwise operations by 25.1X---59.5X
(Section~\ref{sec:lte-analysis}). In addition to these benefits,
Buddy frees up significant processing capacity, cache resources,
and memory bandwidth for other co-running applications, thereby
reducing both computation-unit and memory interference caused by
bulk bitwise operations and thus enabling better {\em overall}
system performance.

\addtocounter{figure}{1}
\begin{figure*}[b]
  \centering
  \begin{minipage}{0.18\textwidth}
  \includegraphics[scale=0.75]{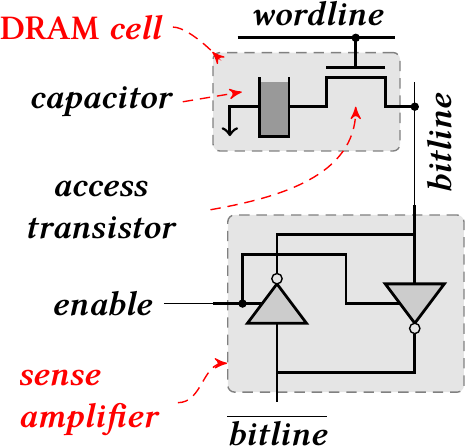}
  \caption{DRAM cell and sense amplifier}
  \label{fig:cell-sa}
  \end{minipage}\qquad\quad
  \begin{minipage}{0.7\textwidth}
  \includegraphics[scale=0.75]{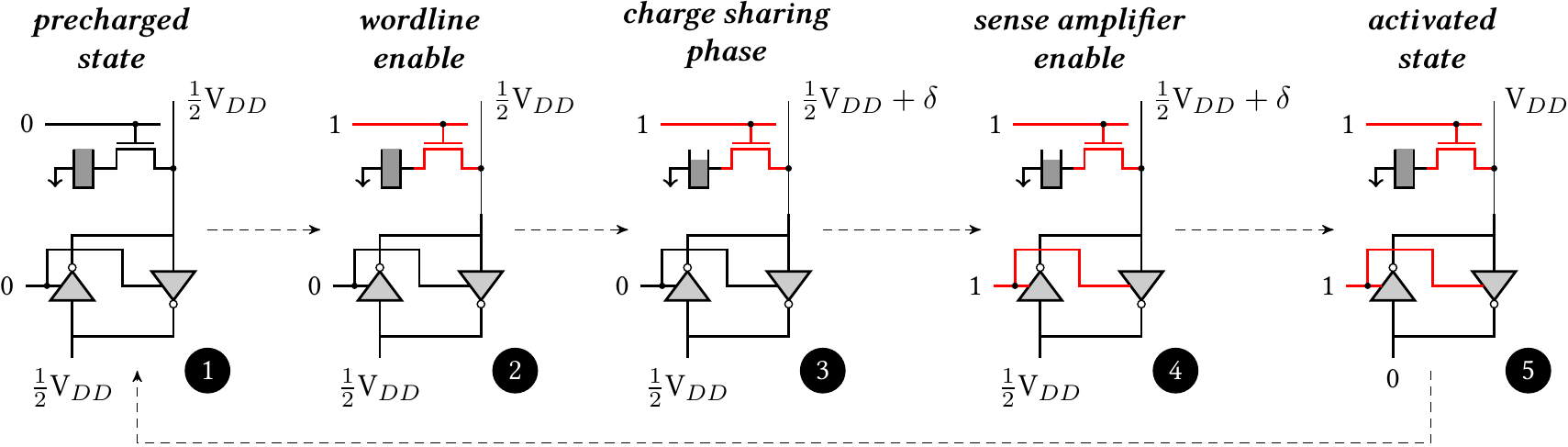}
  \caption{State transitions involved in DRAM cell activation}
  \label{fig:cell-operation}
  \end{minipage}
\end{figure*}
\addtocounter{figure}{-3}

We evaluate three data-intensive applications to demonstrate
Buddy's benefits in comparison to a state-of-the-art baseline
processor that performs bitwise operations using SIMD
extensions. First, we show that Buddy improves end-to-end
performance of queries performed on database bitmap indices by
6.0X, on average across a range of query parameters. Second, Buddy
improves the performance of BitWeaving~\cite{bitweaving}, a
recently proposed technique to accelerate column scan operations
in databases, by 7.0X, on average across a range of scan
parameters. Third, for the commonly-used \emph{set} data
structure, Buddy improves performance of set intersection, union,
and difference operations by 3.0X compared to conventional
implementations~\cite{red-black-tree}.
Section~\ref{sec:applications} describes our simulation
framework~\cite{gem5}, workloads, results, and four other
potential use cases for Buddy: bulk masked initialization,
encryption algorithms, DNA sequence mapping, and approximate
statistics.

We make the following contributions.
\begin{itemize}[leftmargin=*,topsep=2pt]\itemsep0pt\parskip1pt

\item To our knowledge, this is the first work that proposes a
  low-cost mechanism to perform bulk bitwise operations completely
  within a DRAM chip. We introduce Buddy, a mechanism that
  exploits the analog operation of DRAM to perform any row-wide
  bitwise operation efficiently using DRAM.\footnote{The novelty
    of our approach is confirmed by at least one cutting-edge DRAM
    design team in industry~\cite{uksong-kang}. The closest work
    we know of are patents by Mikamonu~\cite{mikamonu}. The
    mechanisms presented in these patents are significantly
    different and costlier than our approach. We discuss this work
    in Section~\ref{sec:related}.}

\item We present a low-cost implementation of Buddy, which
  requires modest changes to the DRAM architecture. We verify our
  implementation of Buddy with rigorous SPICE simulations. The
  cost of our implementation is 1\% of the DRAM chip area, and our
  implementation requires no new DRAM
  commands. (Section~\ref{sec:implementation})

\item We evaluate the benefits of Buddy on both 1) raw
  throughput/energy of seven commonly-used bulk bitwise operations
  and 2) three data-intensive real workloads that make heavy use
  of such operations. Our extensive results show that Buddy
  significantly outperforms the state-of-the-art approach of
  performing such operations in the SIMD units of a CPU or in the
  execution units of a GPU. We show that the large improvements in
  raw throughput of bitwise operations translate into large
  improvements (3.0X-7.0X) in the performance of three
  data-intensive workloads. (Section~\ref{sec:applications})

\end{itemize}

%% file: background.tex
\section{Background on DRAM Operation}
\label{sec:background}

\begin{figure}[h]
  \centering
  \includegraphics[scale=0.7]{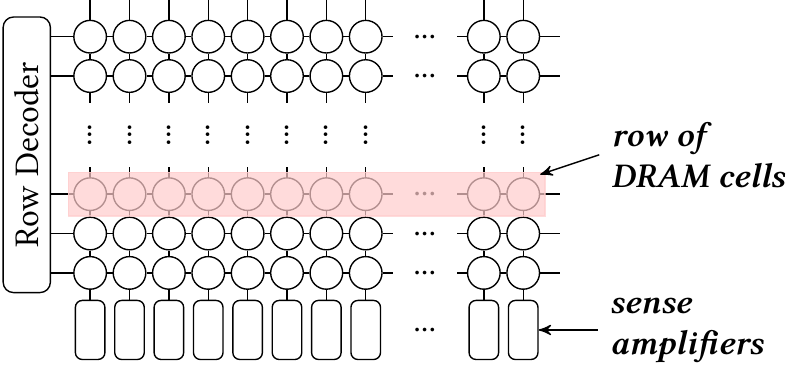}
  \caption{DRAM subarray}
  \label{fig:subarray}
\end{figure}

DRAM-based memory consists of a hierarchy of structures with
\emph{channels}, \emph{modules}, and \emph{ranks} at the top.
Each rank consists of a number of chips that operate in
unison. Hence, a rank can be logically viewed as a single wide
DRAM chip. Each rank is further divided into many
\emph{banks}. All access-related commands are directed towards a
specific bank. Each bank consists of several subarrays and
peripheral logic to process
commands~\cite{salp,dsarp,half-dram,rowclone}. Each subarray
consists of many rows (typically 512/1024) of DRAM cells, a row of
\emph{sense amplifiers}, and a \emph{row address
  decoder}. Figure~\ref{fig:subarray} shows the logical
organization of a subarray.

\addtocounter{figure}{2}

At a high level, accessing data from a subarray involves three
steps. The first step, called \emph{row activation}, copies data
from a specified row of DRAM cells to the corresponding row of
sense amplifiers. This step is triggered by the \cmdact
command. Then, data is accessed from the sense amplifiers using a
\cmdrd or \cmdwr command. Each \cmdrd or \cmdwr accesses only a
subset of sense amplifiers. Once a row is activated, multiple
\cmdrd and \cmdwr commands can be issued to that row. The bank is
then prepared for a new row access by performing an operation
called \emph{precharging}. This step is triggered by a \cmdpre
command. We now explain these operations in detail by focusing our
attention on a single DRAM cell and a sense amplifier.

\subsection{DRAM Cell and Sense Amplifier}

Figure~\ref{fig:cell-sa} shows the connection between a DRAM cell
and a sense amplifier. Each DRAM cell consists of 1)~a
\emph{capacitor}, and 2)~an \emph{access transistor} that controls
access to the cell.  Each sense amplifier consists of two
inverters, and an \emph{enable} signal. The output of each
inverter is connected to the input of the other inverter.  The
wire that connects the cell to the sense amplifier is called the
\emph{bitline}, and the wire that controls the access transistor
is called the \emph{wordline}. We refer to the wire on the other
end of the sense amplifier as \bbar (``bitline bar'').

\subsection{DRAM Cell Operation}

Figure~\ref{fig:cell-operation} shows the state transitions
involved in extracting the state of the DRAM cell. In this figure,
we assume that the cell is initially in the charged state. The
operation is similar if the cell is initially empty. In the
initial \emph{precharged} state \ding{202}, both the bitline and
\bbar are maintained at a voltage level of \halfvdd. The sense
amplifier and the wordline are disabled.

The \cmdact command triggers an access to the cell. Upon receiving
the \cmdact, the wordline of the cell is raised \ding{203},
connecting the cell to the bitline. Since the capacitor is fully
charged, and therefore at a higher voltage level than the bitline,
charge flows from the capacitor to the bitline until both the
capacitor and the bitline reach the same voltage level \halfvddpd
\ding{204}. This phase is called \emph{charge sharing}. After the
charge sharing is complete, the sense amplifier is enabled
\ding{205}. The sense amplifier detects the deviation in the
voltage level of the bitline (by comparing it with the voltage
level on the \bbar). The sense amplifier then amplifies the
deviation to the stable state where the bitline is at the voltage
level of \vdd (and the \bbar is at 0). Since the capacitor is
still connected to the bitline, the capacitor also gets fully
charged \ding{206}. If the capacitor was initially empty, then the
deviation on the bitline would be negative (towards $0$), and the
sense amplifier would drive the bitline to $0$. Each \cmdact
command operates on an entire row of cells (typically 8~KB of data
across a rank).

After the cell is activated, data can be accessed from the bitline
by issuing a \cmdrd or \cmdwr to the column containing the cell
(not shown in Figure~\ref{fig:cell-operation}).  Once the data is
accessed, the subarray needs to be taken back to the initial
precharged state \ding{202}. This is done by issuing a \cmdpre
command. Upon receiving this command, DRAM first lowers the raised
wordline, thereby disconnecting the capacitor from the
bitline. After this, the sense amplifier is disabled, and both the
bitline and the \bbar are driven to the voltage level of \halfvdd.

Our mechanism, Buddy, exploits several aspects of modern DRAM
operation to efficiently perform bitwise operations completely
inside DRAM with low cost. In the following sections, we describe
the two components of Buddy in detail.

%% file: mechanism.tex
\section{\buddyao}
\label{sec:bitwise-and-or}

 The first component of Buddy to perform bitwise AND and OR
 operations exploits two facts about the DRAM operation.
\begin{enumerate}[topsep=2pt]\itemsep0pt\parskip0pt
\item Within a DRAM subarray, each sense amplifier is shared by
  many DRAM cells (typically 512 or 1024).
\item The final state of the bitline after sense amplification
  depends primarily on the voltage deviation on the bitline after
  the charge sharing phase.
\end{enumerate}
Based on these facts, we observe that simultaneously activating
three cells, rather than a single cell, results in a majority
function---i.e., at least two cells have to be fully charged for
the final state to be a logical ``1''. We refer to such a
simultaneous activation of three cells (or rows) as
\emph{triple-row activation}. We now conceptually describe
triple-row activation and how we use it to perform bitwise AND and
OR operations.

\subsection{Triple-Row Activation (\tra)}
\label{sec:triple-row-activation}

A triple-row activation (\tra) simultaneously connects each sense
amplifier with three DRAM cells. For ease of conceptual
understanding, let us assume that all cells have the same
capacitance, the transistors and bitlines behave ideally (no
resistance), and the cells start at a fully refreshed state. Then,
based on charge sharing principles~\cite{dram-cd}, the bitline
deviation at the end of the charge sharing phase of the \tra is,
\begingroup\makeatletter\def\f@size{8}\check@mathfonts\vspace{-2mm}
\begin{eqnarray}
  \delta &=& \frac{k.C_c.V_{DD} + C_b.\frac{1}{2}V_{DD}}{3C_c + C_b}
  - \frac{1}{2}V_{DD} \nonumber \\
  &=& \frac{(2k - 3)C_c}{6C_c + 2C_b}V_{DD} \label{eqn:delta}
\end{eqnarray}
\endgroup
where, $\delta$ is the bitline deviation, $C_c$ is the cell
capacitance, $C_b$ is the bitline capacitance, and $k$ is the
number of cells in the fully charged state. It is clear that
$\delta > 0$ if and only if $2k - 3 > 0$. In other words, the
bitline deviation will be positive if $k = 2, 3$ and it will be
negative if $k = 0, 1$. Therefore, we expect the final state of
the bitline to be \vdd if at least two of the three cells are
initially full charged, and the final state to be $0$, if at least
two of the three cells are initially fully empty.

\begin{figure}[h]
  \centering
  \includegraphics[scale=0.9]{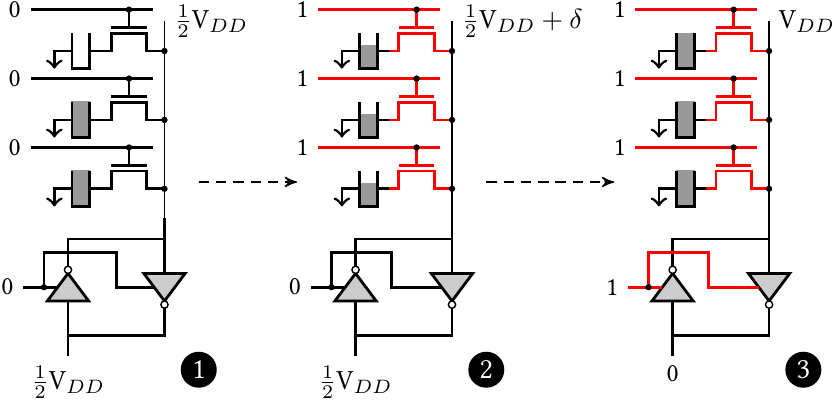}
  \caption{Triple-row activation}
  \label{fig:triple-row-activation}
\end{figure}

Figure~\ref{fig:triple-row-activation} shows an example of \tra
where two of the three cells are initially in the charged state
\ding{202}. When the wordlines of all the three cells are raised
simultaneously \ding{203}, charge sharing results in a positive
deviation on the bitline. Therefore, after sense amplification
\ding{204}, the sense amplifier drives the bitline to \vdd, and as
a result, fully charges all the three cells.

If $A$, $B$, and $C$ represent the logical values of the three
cells, then the final state of the bitline is $AB + BC + CA$ (the
majority function). Importantly, we can rewrite this expression
as $C(A + B) + \overline{C}(AB)$. In other words, by controlling
the value of the cell $C$, we can use \tra to execute a bitwise
AND or bitwise OR of the cells $A$ and $B$. Due to the regular
bulk operation of cells in DRAM, this approach naturally extends
to an entire row of DRAM cells and sense amplifiers, enabling a
multi-kilobyte-wide bitwise AND/OR operation.

\subsection{Making \tra Work}
\label{sec:and-or-challenges}

There are five issues with \tra that we need to resolve for it to
be implementable in a real design.
\begin{enumerate}[labelindent=0pt,topsep=4pt,leftmargin=*]\itemsep0pt\parskip0pt
  \item The deviation on the bitline with three cells may not be
    large enough to be detected by the sense amplifier or it could
    lengthen the sense amplification process.
  \item Equation~\ref{eqn:delta} assumes that all cells have the
    same capacitance. However, due to process variation, different
    cells will have different capacitance. This can affect the
    reliability of \tra.
  \item As shown in Figure~\ref{fig:triple-row-activation}
    (state~\ding{204}), the \tra overwrites the data of all the
    three cells with the final value.  In other words, \tra
    modifies all source data.
  \item Equation~\ref{eqn:delta} assumes that the cells involved
    in a \tra are either fully charged or fully empty. However,
    DRAM cells leak charge over time. Therefore, \tra may not
    operate as expected, if the cells involved had leaked charge.
  \item Simultaneously activating three \emph{arbitrary} rows
    inside a DRAM subarray requires the memory controller to
    communicate three row addresses to the DRAM module and the
    subarray row decoder to simultaneously decode three row
    addresses. This introduces an enormous cost on the address bus
    and the row decoder.
\end{enumerate}

We address the first two issues by performing rigorous and
detailed SPICE simulations of \tra. The results of these confirm
that \tra works as expected even with $\pm 20\%$ process
variation. Section~\ref{sec:spice-sim} presents these results.  We
propose a simple implementation of \buddyao that addresses
\emph{all} of the last three issues. We describe this
implementation in
Section~\ref{sec:and-or-mechanism}. Section~\ref{sec:implementation}
describes the final implementation of Buddy.

\begin{figure*}[b]
  \centering
  \begin{minipage}{0.2\textwidth}
  \includegraphics[scale=0.75]{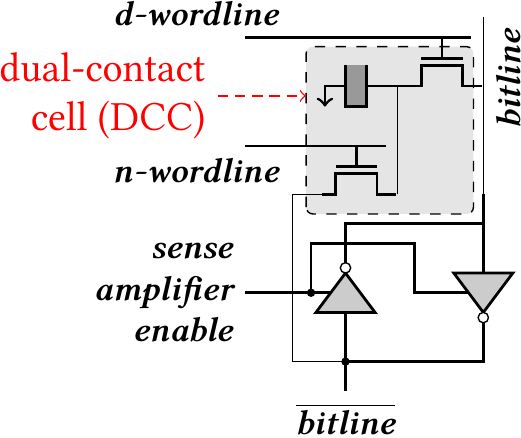}
  \caption{A dual-contact cell connected to both ends of a sense
    amplifier}
  \label{fig:dcc-not}
  \end{minipage}\qquad\quad
  \begin{minipage}{0.7\textwidth}
  \includegraphics[scale=0.9]{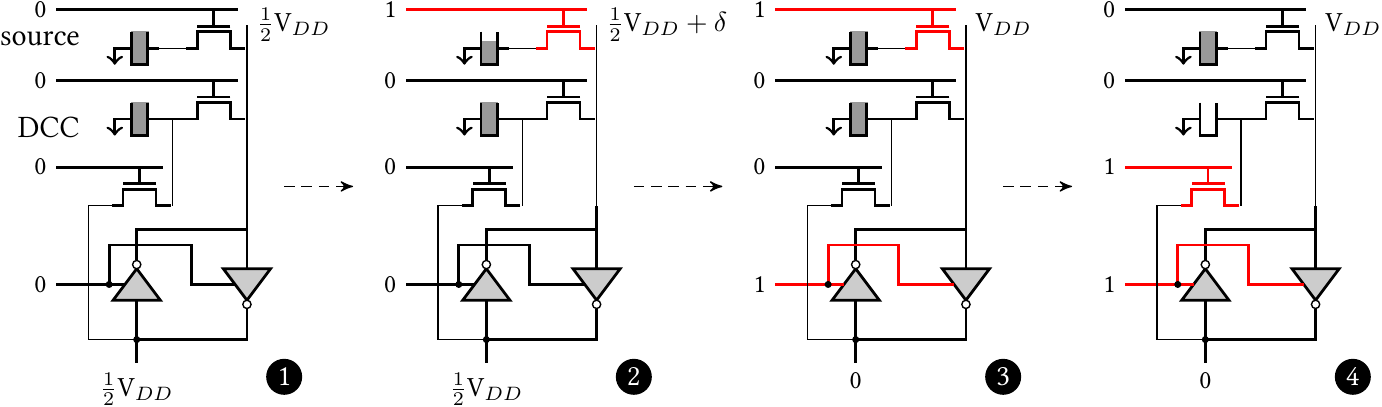}
  \caption{Bitwise NOT using a dual contact capacitor}
  \label{fig:bitwise-not}
  \end{minipage}
\end{figure*}

\subsection{SPICE Simulations}
\label{sec:spice-sim}

We perform SPICE simulations to confirm the reliability of
\tra. We implement the DRAM sense amplifier circuit using 55nm
DDR3 model parameters~\cite{rambus}, and PTM low-power transistor
models~\cite{ptmweb,zhaoptm}. We use cell/transistor parameters
from the Rambus power model~\cite{rambus} (cell capacitance =
22fF, transistor width = 55nm, transistor height = 85nm).

The DRAM specification is designed for the worst case conditions
when a cell has not been accessed for a long time (refresh
interval). Therefore, we can activate a fully refreshed cell with
a latency much lower than the standard activation latency,
35ns~\cite{charge-cache,nuat}. In fact, we observe that we can
activate a fully charged/empty cell in 20.9~ns/13.5~ns.

For \tra, there are four possible cases depending on the number
cells that are initially fully charged. For these cases, we add
different levels of process variation among cells, so that the
strong cell attempts to override the majority decision of the two
weak cells. Table~\ref{tab:pv} shows the latency of \tra for the
four possible cases with different levels of process variation,
where \emph{s} and \emph{w} subscripts stand for \emph{strong}
and \emph{weak}, respectively, for the cells that store either 0
or 1.  

\begin{table}[h]\footnotesize
  \centering
  \input{tables/pv}
  \caption{Effect of process variation on the latency of
    triple-row activation. All times are in ns.}
  \label{tab:pv}
\end{table}

We draw three conclusions. First, for the cases where all three
cells are either 0 or 1, the latency of \tra is stable at around
16~ns and 22~ns respectively, even in the presence of process
variation. This is because, in these two cases, all three cells
push the bitline toward the same direction. Second, for the other
two cases, while the latency of \tra increases with increasing
process variation, it is still well within the DRAM specification
even with $\pm$20\% process variation (i.e., \emph{a 40\%
difference in cell capacitance}). Third, we observe the first
failure at $\pm$25\% for the $1_s0_w0_w$ case. In this case,
the sense amplifier operates incorrectly by detecting a ``1''
instead of ``0''. In summary, our SPICE simulations show that
\tra works as expected even in the presence of significant
process variation.

Prior works~\cite{al-dram,data-retention,vrt-1,vrt-2} show that
temperature increases DRAM cell leakage. However, as we show in
the next section, our mechanism \emph{always} ensures that the
cells involved in the TRA are fully refreshed just before the
operation. Therefore, we do not expect temperature to affect the
reliability of TRA.

\subsection{Implementation of \buddyao}
\label{sec:and-or-mechanism}

To avoid modification of the source data (issue 3), our
implementation reserves a set of \emph{designated rows} in each
subarray that will be used to perform {\tra}s. These designated
rows are chosen statically at \emph{design time}. To perform
bitwise AND or OR operation on two arbitrary sources rows, our
mechanism first copies the data of the source rows into the
designated rows and performs the required \tra on the designated
rows. Our final implementation (Section~\ref{sec:implementation})
reserves four designated rows in each subarray
(\taddr{0}---\taddr{3}). As an example, to perform a bitwise AND
of two rows \texttt{A} and \texttt{B}, and store the result in row
\texttt{R}, our mechanism performs the following
steps.\vspace{-1mm}
\begin{enumerate}[topsep=4pt]\itemsep0pt\parsep0pt\parskip0pt\small
\item \emph{Copy} data of row \texttt{A} to row \taddr{0}
\item \emph{Copy} data of row \texttt{B} to row \taddr{1}
\item \emph{Initialize} row \taddr{2} to $0$
\item \emph{Activate} rows \taddr{0}, \taddr{1}, and \taddr{2} simultaneously
\item \emph{Copy} data of row \taddr{0} to row \texttt{R}
\end{enumerate}\vspace{-1mm}

This implementation allows us to address the last three issues
described in Section~\ref{sec:and-or-challenges}.  First, by not
performing the \tra directly on the source data, our mechanism
trivially avoids modification of the source data (issue
3). Second, each copy operation takes five orders of magnitude
lower latency (1~${\mu}s$) than the refresh interval
(64~ms). Since these copy operations are performed \emph{just
  before} the \tra, the rows involved in the \tra are very close
to the fully refreshed state just before the operation (addressing
issue 4). Finally, since the \emph{designated} rows are chosen at
design time, the memory controller can use a reserved address to
communicate \tra of a \emph{pre-defined} set of three designated
rows. To this end, our mechanism reserves a set of row addresses
\emph{just} to control the designated rows. While some of these
addresses perform single row activation of the designated rows
(necessary for the copy operations), others trigger {\tra}s of
pre-defined sets of designated rows. For instance, in our final
implementation (Section~\ref{sec:implementation}), to perform a
\tra of designated rows \taddr{0}, \taddr{1}, and \taddr{2} (step
4, above), the memory controller simply issues a \cmdact with the
reserved address \baddr{12}. The row decoder maps \baddr{12} to
all the three wordlines of rows \taddr{0}, \taddr{1}, and
\taddr{2}. This mechanism requires \emph{no} changes to the
address bus and significantly reduces the cost and complexity of
the row decoder compared to performing \tra on three
\emph{arbitrary} rows (addressing issue 5).

\subsection{Mitigating the Overhead of Copy Operations}
\label{sec:and-or-rowclone}

Our mechanism needs a set of copy and initialization operations to
copy the source data into the designated rows and copy the result
back to the destination. These copy operations, if performed
naively, will nullify the benefits of our mechanism. Fortunately,
a recent work, RowClone~\cite{rowclone}, has proposed two
techniques to copy data between two rows quickly and efficiently
within DRAM. The first technique, RowClone-FPM (Fast Parallel
Mode), copies data within a subarray by issuing two back-to-back
{\cmdact}s to the source row and the destination row. The second
technique, RowClone-PSM (Pipelined Serial Mode), copies data
between two banks by using the shared internal bus to overlap the
read to the source bank with the write to the destination bank.

With RowClone, we can perform all the copy operations and the
initialization operation efficiently within DRAM. To use RowClone
for the initialization operation, we reserve two additional
\emph{control} rows, \czero and \cone. \czero is pre-initialized
to $0$ and \cone is pre-initialized to 1. Depending on the
operation to be performed, our mechanism uses RowClone to copy
either \czero or \cone to the appropriate designated row.

In the best case, when all the three rows involved in the
operation are in the same subarray, our mechanism uses
RowClone-FPM for all copy and initialization operations. However,
if the three rows are in different subarrays, some of the three
copy operations have to use RowClone-PSM. In the worst case, when
all three copy operations have to use RowClone-PSM, our approach
would consume higher latency than the baseline. However, when only
one or two RowClone-PSM operations are required, our mechanism is
faster and more energy-efficient than existing systems.

\section{\buddynot}
\label{sec:bitwise-not}

\buddynot exploits the following fact that at the end of the sense
amplification process, the voltage level of the \bbar contains the
negation of the logical value of the cell.  Our key idea to
perform bitwise NOT in DRAM is to transfer the data on the \bbar
to a cell that can be connected to the bitline. For this purpose,
we introduce the \emph{dual-contact cell}. A dual-contact cell
(DCC) is a DRAM cell with two transistors (a 2T-1C cell similar to
the one described in~\cite{2t-1c-1}).  For each DCC, one
transistor connects the DCC to the bitline and the other
transistor connects the DCC to the \bbar.  We refer to the
wordline that controls the connection between the DCC and the
bitline as the \dwordline (or data wordline). We refer to the
wordline that controls the connection between the DCC and the
\bbar as the \nwordline (or negation
wordline). Figure~\ref{fig:dcc-not} shows a DCC connected to a
sense amplifier.

Figure~\ref{fig:bitwise-not} shows the steps involved in
transferring the negation of a \emph{source} cell on to the DCC
connected to the same sense amplifier \ding{202}. Our mechanism
first activates the source cell \ding{203}. The activation process
drives the bitline to the data corresponding to the source cell,
\vdd in this case \ding{204}. More importantly, for the purpose of
our mechanism, it drives the \bbar to $0$. In this state, our
mechanism activates the \emph{n-wordline}, enabling the transistor
that connects the DCC to the \bbar~\ding{205}. Since the \bbar is
already at a stable voltage level of $0$, it overwrites the value
in the DCC with $0$, essentially copying the negation of the
source data into the DCC. After this step, we can efficiently copy
the negated data into the destination cell using RowClone.

\textbf{SPICE Simulations.} We perform detailed SPICE simulation
of the dual-contact cell with the same cell and transistor
parameters described in Section~\ref{sec:spice-sim}. Our
simulation results confirm that the DCC operation described in
Figure~\ref{fig:bitwise-not} works as expected. We do not present
details due to lack of space.

\textbf{Implementation of \buddynot.} Our implementation adds two
rows of DCCs to each subarray, one on each side of the row of
sense amplifiers. Similar to the designated rows used for \buddyao
(Section~\ref{sec:and-or-mechanism}), the memory controller uses
reserved row addresses to control the \emph{d-wordline}s and
\emph{n-wordline}s of the DCC rows---e.g., in our final
implementation (Section~\ref{sec:implementation}), address
\baddr{5} maps to the \emph{n-wordline} of the first DCC row. To
perform a bitwise NOT of row \texttt{A} and store the result in
row \texttt{R}, the memory controller performs the following
steps.

\begin{enumerate}[topsep=2pt]\itemsep0pt\parsep0pt\parskip0pt\small
\item \emph{Activate} row \texttt{A}
\item \emph{Activate} \emph{n-wordline} of DCC (address \baddr{5})
\item \emph{Precharge} the bank.
\item \emph{Copy} data from \emph{d-wordline} of DCC to row \texttt{R}
\end{enumerate}\vspace{-1mm}

Similar to the copy operations in \buddyao
(Section~\ref{sec:and-or-rowclone}), the copy operation in Step 4
above be efficiently performed using RowClone.

%% file: tables/pv.tex
\newcommand{\pvar}[1]{$\pm$#1\%}
\begin{tabular}{rcccccc}
  \toprule
  Variation & \pvar{0} & \pvar{5} & \pvar{10} & \pvar{15} &
  \pvar{20} & \pvar{25}\\
  \midrule
  $0_s 0_w 0_w$ & 16.4 & 16.3 & 16.3 & 16.4 & 16.3 & 16.2\\
  $1_s 0_w 0_w$ & 18.3 & 18.6 & 18.8 & 19.1 & 19.7 & \textbf{Fail}\\
  $0_s 1_w 1_w$ & 24.9 & 25.0 & 25.2 & 25.3 & 25.4 & 25.7\\
  $1_s 1_w 1_w$ & 22.5 & 22.3 & 22.2 & 22.2 & 22.2 & 22.1\\
  \bottomrule
\end{tabular}

%% file: implementation.tex
\section{Buddy: Putting It All Together}
\label{sec:implementation}

In this section, we describe how we integrate \buddyao and
\buddynot into a single mechanism that can perform any bitwise
operation efficiently inside DRAM. First, both \buddyao and
\buddynot reserve a set of rows in each subarray and a set of
addresses that map to these rows. We present the final set of
reserved addresses and their mapping in detail
(Section~\ref{sec:address-grouping}). Second, we introduce a new
primitive called \aap (\cmdact-\cmdact-\cmdpre) that the memory
controller uses to execute various bitwise operations
(Section~\ref{sec:command-sequence}). Finally, we describe an
optimization that lowers the latency of the \aap primitive,
further improving the performance of Buddy
(Section~\ref{sec:split-row-decoder}).

\subsection{Row Address Grouping}
\label{sec:address-grouping}

\begin{figure}[h]
  \centering
  \includegraphics[scale=0.9]{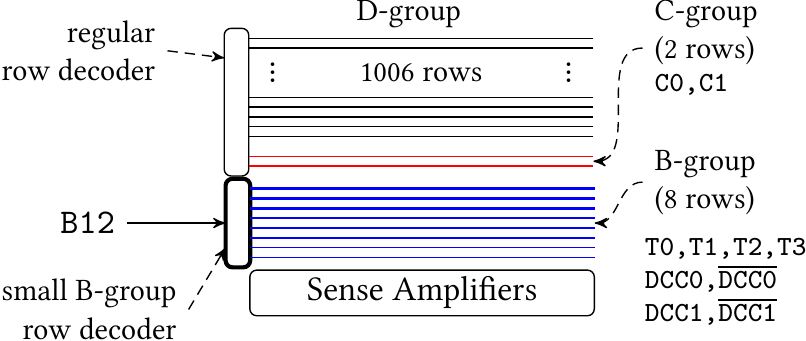}
  \caption{Row address grouping. The figure shows how the B-group
    row decoder (Section~\ref{sec:split-row-decoder})
    simultaneously activates rows \taddr{0}, \taddr{1}, and
    \taddr{2} with a single address \baddr{12}.}
  \label{fig:row-address-grouping}
\end{figure}

Our implementation divides the space of row addresses in each
subarray into three distinct groups
(Figure~\ref{fig:row-address-grouping}): 1)~bitwise group,
2)~control group, and 3)~data group.

The \emph{B-group} (or the \emph{bitwise} group) corresponds to
the addresses used to perform the bitwise operations. This group
contains eight physical wordlines: four corresponding to the
\emph{designated} rows (\taddr{0}---\taddr{3}) used to perform
triple-row activations (Section~\ref{sec:and-or-mechanism}) and
the remaining four corresponding to the
\emph{d-}and-\emph{n-wordlines} that control the two rows of
dual-contact cells (Section~\ref{sec:bitwise-not}). We refer to
the {\dwordline}s of the two rows as \dccdz and \dccdo, and the
corresponding {\nwordline}s as \dccnz and \dccno. The B-group
contains 16 reserved addresses:
\baddr{0}---\baddr{15}. Table~\ref{table:b-group-mapping} lists
the mapping between the 16 addresses and the wordlines. The first
eight addresses individually activate each of the 8 wordlines in
the group. Addresses \baddr{12}---\baddr{15} activate three
wordlines simultaneously. Buddy uses these addresses to trigger
triple-row activations. Finally, addresses \baddr{8}---\baddr{11}
activate two wordlines. As we will show in the next section, Buddy
uses these addresses to copy the result of an operation
simultaneously to two rows (e.g., zero out two rows
simultaneously).\footnote{An implementation can reserve more rows
  for the B-group. While this will reduce the number of rows
  available to store application data, it can potentially reduce
  the number of copy operations required to implement different
  sequences of bitwise operations.}

\begin{table}[h]\small
  \centering
  \input{tables/b-group-mapping}
  \caption{Mapping of B-group addresses}
  \label{table:b-group-mapping}
\end{table}

The \emph{C-group} (or the \emph{control} group) contains the two
pre-initialized rows for controlling the bitwise AND/OR operations
(Section~\ref{sec:and-or-rowclone}). Specifically, this group
contains two addresses: \czero (all zeros) and \cone (all ones).

The \emph{D-group} (or the \emph{data} group) corresponds to the
rows that store regular data. This group contains all the
addresses that are neither in the \emph{B-group} nor in the
\emph{C-group}. Specifically, if each subarray contains $1024$
rows, then the \emph{D-group} contains $1006$ addresses, labeled
\daddr{0}---\daddr{1005}. Buddy exposes only the D-group addresses
to the operating system (OS). To ensure that the OS system has a
contiguous view of memory, the memory controller interleaves the
row addresses of subarrays such that the D-group addresses across
all subarrays are mapped contiguously to the physical address
space.

With these address groups, the memory controller can use the
\emph{existing} command interface to communicate all variants of
\cmdact to the DRAM chips. Depending on the address group, the
DRAM chips internally process the \cmdact appropriately.  For
instance, by just issuing an \cmdact to address \baddr{12}, the
memory controller can simultaneously activate rows \taddr{0},
\taddr{1}, and \taddr{2}. We will now describe how the memory
controller uses this interface to express bitwise operations.

\subsection{Executing Bitwise Ops: The AAP Primitive}
\label{sec:command-sequence}
Let us consider the operation, \daddr{k} \texttt{=} \bnot
\daddr{i}. To perform this bitwise NOT operation, the memory
controller sends the following sequence of commands.

{\tabcolsep4pt\small
\begin{tabular}{llllll}
1. & \cmdact \daddr{i}; & 2. & \cmdact \baddr{5}; & 3. & \cmdpre;\\
4. & \cmdact \baddr{4}; & 5. & \cmdact \daddr{k}; & 6. & \cmdpre;
\end{tabular}
}

The first three steps are the same as those described in
Section~\ref{sec:bitwise-not}. These operations essentially copy
the negation of row \daddr{i} into the DCC row 0 (as described in
Figure~\ref{fig:bitwise-not}). Step 4 activates the \dwordline of
the DCC row, transferring the negation of the source data on to
the bitlines. Finally, Step 5 activates the destination row,
copying the data on the bitlines, i.e., the negation of the source
data, to the destination row.

If we observe the negation operation, it consists of two steps of
\cmdact-\cmdact-\cmdpre operations. We refer to this sequence as
the \aap primitive. Each \aap takes two addresses as
input. \texttt{\aap(addr1, addr2)} corresponds to the following
sequence of commands: \texttt{\cmdact addr1; \cmdact addr2;
  \cmdpre;} Logically, an \aap operation copies the result of
activating the first address (\texttt{addr1}) to the row mapped to
the second address (\texttt{addr2}).

\begin{figure}[h]
  \includegraphics[scale=0.84]{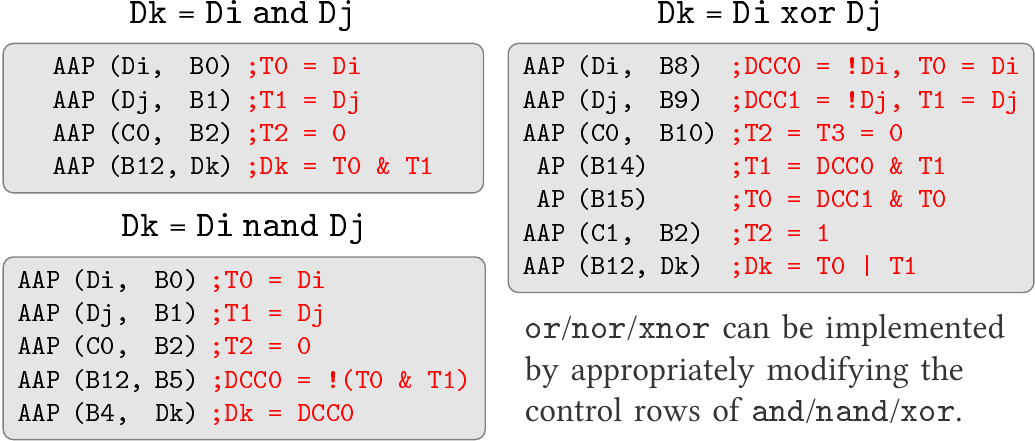}
  \caption{Command sequences for different bitwise operations}
  \label{fig:command-sequences}
\end{figure}

We observe that most bitwise operations mainly involve a sequence
of \aap operations. In a few cases, they require a regular \cmdact
followed by a \cmdpre, which refer to as \ap.  \ap takes one
address as input. \texttt{\ap(addr)} corresponds to the following
commands: \texttt{\cmdact addr; \cmdpre;}
Figure~\ref{fig:command-sequences} shows the sequence of steps
taken by the memory controller to execute three bitwise
operations: \band, \bnand, and \bxor.

Let us consider the \band operation, \daddr{k} = \daddr{i} \band
\daddr{j}. The four \aap operations directly map to the steps
described in Section~\ref{sec:and-or-mechanism}. The first \aap
copies the first source row (\daddr{i}) into the designated row
\taddr{0}. Similarly, the second \aap copies the second source row
\daddr{j} to row \taddr{1}, and the third \aap copies the control
row ``0'' to row \taddr{2} (to perform a bitwise AND). Finally,
the last \aap first issues an \cmdact to address \baddr{12}. As
described in Table~\ref{table:b-group-mapping}, this command
simultaneously activates the rows \taddr{0}, \taddr{1}, and
\taddr{2}, resulting in an \band operation of the values of rows
\taddr{0} and \taddr{1}. This command is immediately followed by
an \cmdact to \daddr{k}, which in effect copies the result of the
\band operation to the destination row \daddr{k}.

While an individual bitwise operation involves multiple copy
operations, this overhead of copying can be reduced by applying
standard compiler techniques. For instance, accumulation-like
operations generate intermediate results that are immediately
consumed. An optimization like dead-store elimination may prevent
these values from being needlessly copied. Our evaluations
(Section~\ref{sec:applications}) consider the overhead of the copy
operations without these optimizations.

\subsection{Accelerating AAP with a Split Row Decoder}
\label{sec:split-row-decoder}

It is clear from Figure~\ref{fig:command-sequences} that the
latency of executing any bitwise operation using Buddy depends on
the latency of the \aap primitive. The latency of the \aap in turn
depends on the latency of \cmdact, i.e., \tras, and the latency of
\cmdpre, i.e., \trp. The naive approach to execute an \aap is to
perform the three operations serially. Using this approach, the
latency of \aap is 2\tras + \trp. While even this naive approach
can offer better throughput and energy efficiency than existing
systems (not shown due to space limitations), we propose an
optimization that significantly reduces the latency of \aap.

Our optimization is based on the following two
observations. First, the second \cmdact of an \aap is issued to an
already \emph{activated bank}. As a result, this \cmdact does not
require full sense amplification, which is the dominant portion of
\tras. Second, if we observe all the bitwise operations in
Figure~\ref{fig:command-sequences}, with the exception of one \aap
in \bnand, \emph{exactly one} of the two \cmdacts in each \aap is
to a \emph{B-group} address.

To exploit these observations, our mechanism splits the row
decoder into two parts. The first part decodes all
\emph{C/D-group} addresses and the second smaller part decodes
\emph{only} \emph{B-group} addresses. Such a split allows the
subarray to \emph{simultaneously} decode a \emph{C/D-group}
address along with a \emph{B-group} address. With this setup, if
the memory controller issues the second \cmdact of an \aap after
the first activation has sufficiently progressed, the sense
amplifier will force the data of the second row(s) to the result
of the first activation. This mechanism allows the memory
controller to significantly overlap the latency of the two
\cmdacts. This approach is similar to the inter-segment copy
operation used by Tiered-Latency DRAM~\cite{tl-dram}. Based on
SPICE simulations, our estimate of the latency of executing both
the {\cmdact}s is 4~ns larger than \tras. For DDR3-1600 (8-8-8)
timing parameters~\cite{ddr3-1600}, this optimization reduces the
latency of \aap from 80~ns to 49~ns.

In addition to reducing the latency of \aap, the split row decoder
significantly reduces the complexity of the row decoding
logic. Since only addresses in the \emph{B-group} are involved in
triple-row activations, the complexity of simultaneously raising
three wordlines is restricted to the small \emph{B-group} decoder.

\subsection{DRAM Chip and Controller Cost}
\label{sec:hardware-cost}

Buddy has three main sources of cost to the DRAM chip. First, it
requires the row decoding logic to distinguish between the
\emph{B-group} addresses and the remaining addresses. Within the
\emph{B-group}, it must implement the mapping described in
Table~\ref{table:b-group-mapping}. As the \emph{B-group} contains
only 16 addresses, the complexity of the changes to the row
decoding logic are low.  The second source of cost is the
implementation of the dual-contact cells (DCCs). In our design,
each sense amplifier has only one DCC on each side, and each DCC
has two wordlines associated with it. This design is similar to
the one described in~\cite{2t-1c-1}. In terms of area, the cost of
each DCC is roughly equivalent to two DRAM cells.  The third
source of cost is the capacity lost due to the reserved rows in
the \emph{B-group} and \emph{C-group}.  The system cannot use
these rows to store application data. Our proposed implementation
of Buddy reserves 10 rows in each subarray for the two groups. For
a typical subarray size of 1024 rows, the loss in memory capacity
is $\approx$1\%.

DRAM manufacturers have to test chips to determine if TRA and the
DCCs work as expected. However, since these operations concern
only 8 DRAM rows of the B-group, we expect the additional overhead
of testing to be low.

On the controller side, Buddy requires the memory controller to
1)~store information about different address groups, 2)~track the
timing for different variants of the \cmdact (with or without the
optimizations), and 3)~track the status of different on-going
bitwise operations. While scheduling different requests, the
controller 1)~adheres to power constraints like tFAW, and 2)~can
interleave the multiple AAP commands to perform a bitwise
operation with other requests from different applications. We
believe this modest increase in the DRAM chip/controller
complexity is negligible compared to the improvement in throughput
and energy enabled by Buddy (described in
Sections~\ref{sec:lte-analysis} and~\ref{sec:applications}).

%% file: tables/b-group-mapping.tex
\begin{tabular}{rl}
  \toprule
  \textbf{Addr.} & \textbf{Wordline(s)}\\
  \toprule
  \baddr{0} & \taddr{0}\\
  \baddr{1} & \taddr{1}\\
  \baddr{2} & \taddr{2}\\
  \baddr{3} & \taddr{3}\\
  \baddr{4} & \dccdz\\
  \baddr{5} & \dccnz\\
  \baddr{6} & \dccdo\\
  \baddr{7} & \dccno\\
  \bottomrule
\end{tabular}\quad
\begin{tabular}{rl}
  \toprule
  \textbf{Addr.} & \textbf{Wordline(s)}\\
  \toprule
  \baddr{8} & \dccnz, \taddr{0}\\
  \baddr{9} & \dccno, \taddr{1}\\
  \baddr{10} & \taddr{2}, \taddr{3}\\
  \baddr{11} & \taddr{0}, \taddr{3}\\
  \baddr{12} & \taddr{0}, \taddr{1}, \taddr{2}\\
  \baddr{13} & \taddr{1}, \taddr{2}, \taddr{3}\\
  \baddr{14} & \dccdz, \taddr{1}, \taddr{2}\\
  \baddr{15} & \dccdo, \taddr{0}, \taddr{3}\\
  \bottomrule
\end{tabular}

%% file: support.tex
\section{Integrating Buddy with the System Stack}
\label{sec:support}

We envision two distinct ways to integrate Buddy into the
system. The first way is a loose integration, where we treat Buddy
as an accelerator like GPU. The second way is a tight integration,
where we enable ISA support to integrate Buddy inside main
memory. We now discuss both these ways.

\subsection{Buddy as an Accelerator}

In this approach, the manufacturer designs Buddy as an accelerator
that can be plugged into the system as a separate device. We
envision a system wherein the data structure that relies heavily
on bitwise operations is designed to fit inside the accelerator
memory, thereby minimizing communication between the CPU and the
accelerator. In addition to the performance benefits of
accelerating bitwise operations, this approach has two further
benefits. First, a \emph{single} manufacturer designs both the
DRAM and the memory controller (not true of commodity
DRAM). Second, the details of the data mapping to suit Buddy can
be hidden behind the device driver, which can expose a
simple-to-use API to the applications. Both these factors simplify
the implementation. The execution model for this approach is
similar to that of a GPU, wherein the programmer explicitly
specifies the portions of the program that have to be executed in
the Buddy accelerator.

\subsection{Integrating Buddy with System Main Memory}

A tighter integration of Buddy with the system main memory
requires support from different layers of the system stack, which
we discuss below.

\subsubsection{ISA Support.}

To enable software to communicate occurrences of bulk bitwise
operations to the processor, we introduce new instructions of the
form,\\ \centerline{\texttt{bop dst, src1, [src2], size}}

where \texttt{bop} is the bitwise operation to be performed,
\texttt{dst} is the destination address, \texttt{src1} and
\texttt{src2} are the source addresses, and \texttt{size} denotes
the length of operation in bytes.

\subsubsection{Implementing the New Buddy Instructions.}

Since all Buddy operations are row-wide, Buddy requires the source
and destination rows to be row-aligned and the operation to be at
least the size of a DRAM row. The microarchitecture implementation
of the \texttt{bop} instructions checks if each instance of these
instructions satisfies this constraint. If so, the CPU sends the
operation to the memory controller.  The memory controller in turn
determines the number of RowClone-PSM operations required to
complete the bitwise operation. If the number of RowClone-PSM
operations required is three (in which case performing the
operation using the CPU will be faster,
Section~\ref{sec:and-or-rowclone}), or if the source/destination
rows do not satisfy the alignment/size constraints, the CPU
executes the operation itself. Otherwise, the memory controller
completes the operation using Buddy.  Note that the CPU performs
the virtual-to-physical address translation of the source and
destination rows \emph{before} performing these checks and
exporting the operations to the memory controller. Therefore,
there is no need for any address translation support in the memory
controller.

\subsubsection{Maintaining On-chip Cache Coherence.}

Buddy directly reads/modifies data in main memory. Therefore,
before performing any Buddy operation, the memory controller must
1)~flush any dirty cache lines from the source rows, and
2)~invalidate any cache lines from destination rows. While
flushing the dirty cache lines of the source rows is on the
critical path of a Buddy operation, simple structures like the
Dirty-Block Index~\cite{dbi} can speed up this step. Our mechanism
invalidates the cache lines of the destination rows in parallel
with the Buddy operation. Such a coherence mechanism is already
required by Direct Memory Access (DMA)~\cite{linux-dma}, which is
supported by most modern processors.

\subsubsection{Software Support.}

The minimum support that Buddy requires from software is for the
application to use the new Buddy instructions to communicate the
occurrences of bulk bitwise operations to the processor. However,
as an optimization to enable maximum benefit, the OS can allocate
pages that are likely to be involved in a bitwise operation such
that 1)~they are row-aligned, and 2)~belong to the same
subarray. Note that the OS can still interleave the pages of a
single data structure to multiple subarrays. Implementing this
support requires the OS to be aware of the subarray mapping, i.e.,
determine if two physical pages belong to the same subarray or
not. The OS can extract this information from the DRAM modules
with the help of our memory controller (similar
to~\cite{salp,tl-dram}).


%% file: lte-analysis.tex
\section{Analysis of Throughput \& Energy}
\label{sec:lte-analysis}

\begin{figure}[h]
  \includegraphics{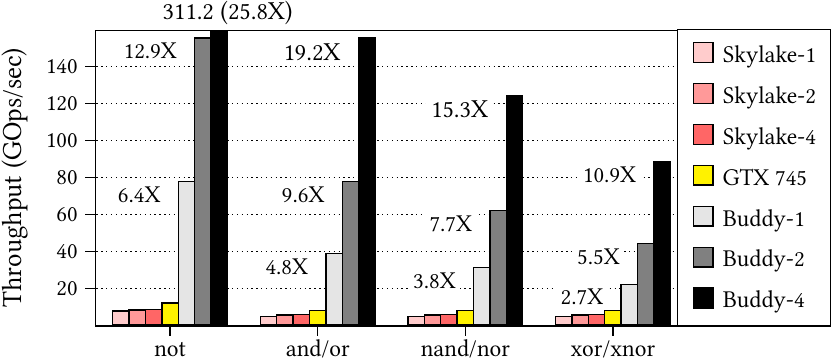}
  \caption{Comparison of throughput of bitwise operations. The
    values on top of each Buddy bar indicates the factor
    improvement in throughput of Buddy on top of the GTX 745.}
  \label{plot:cgb-throughput}
\end{figure}

We compare the throughput of Buddy for bulk bitwise operations to
that of an Intel Skylake Core i7 system~\cite{intel-skylake} and
an NVIDIA GeForce GTX 745 GPU~\cite{gtx745}. The Skylake system
has 4 cores with support for Advanced Vector
eXtensions~\cite{intel-avx}, and two 64-bit DDR3-2133
channels. The GTX 745 contains 3 streaming multi-processors each
with 128 cuda cores. The memory system consists of one 128-bit
DDR3-1800 channel. For each bitwise operation, we run a
microbenchmark that performs the operation repeatedly for many
iterations on large input vectors (32~MB), and measure the
throughput for the operation. Figure~\ref{plot:cgb-throughput}
plots the results of this experiment for six configurations: the
Skylake system with 1, 2, and 4 cores, the GTX 745, and Buddy RAM
with 1, 2, and 4 DRAM banks.

We draw three conclusions. First, for all bitwise operations, the
throughput of the Skylake system is roughly the same for all three
core configurations. We find that the available memory bandwidth
limits the throughput of these operations, and hence using more
cores is not beneficial. Second, while the GTX 745 slightly
outperforms the Skylake system, its throughput is also limited by
the available memory bandwidth. Although a more powerful GPU with
more bandwidth would enable higher throughput, such high-end GPUs
are significantly costlier and also consume very high
energy. Third, even with a single DRAM bank, Buddy significantly
outperforms both the Skylake and the GTX 745 for \emph{all}
bitwise operations (2.7X---6.4X better throughput than the GTX
745). More importantly, unlike the other two systems, Buddy is not
limited by the memory channel bandwidth. Therefore, the throughput
of Buddy scales linearly with increasing number of banks. Even
with power constraints like tFAW, Buddy with two or four banks can
achieve close to an order of magnitude higher throughput than the
other two systems.

We estimate energy for DDR3-1333 using the Rambus power
model~\cite{rambus}. Our energy numbers include only the DRAM and
channel energy, and not the energy consumed by the on-chip
resources. For Buddy, some activate operations have to raise
multiple wordlines and hence consume higher energy. Based on our
analysis, we increase the activation energy by 22\% for each
additional wordline raised.  Table~\ref{table:energy} shows the
energy consumed per kilo byte for different bitwise operations.
Across all bitwise operations, Buddy reduces energy consumption by
at least 25.1X (up to 59.5X) compared to the DDR interface.

\begin{table}[h]\footnotesize
  \centering
  \input{tables/energy}
  \caption{Comparison of energy for various groups of bitwise
    operations. ($\downarrow$) indicates reduction in energy of
    Buddy over the traditional DDR3 interface.}
  \label{table:energy}
\end{table}

Based on these results, we conclude that for systems using
DRAM-based main memory, Buddy is the most efficient way of
performing bulk bitwise operations.

%% file: tables/energy.tex
\setlength{\tabcolsep}{5pt}
\begin{tabular}{ccrrrr}
  \toprule
  & Interface & \bnot & \band/\bor & \bnand/\bnor & \bxor/\bxnor\\
  \toprule
  \multirow{2}{*}{Energy} & DDR3 & 93.7 & 137.9 & 137.9 & 137.9\\
  \multirow{2}{*}{(nJ/KB)} & Buddy & 1.6 & 3.2 & 4.0 & 5.5\\
  &  ($\downarrow$) & 59.5X & 43.9X & 35.1X & 25.1X \\
  \bottomrule  
\end{tabular}

%% file: applications.tex
\section{Effect on Real-World Applications}
\label{sec:applications}

We evaluate the benefits of Buddy on real-world applications using
the Gem5 simulator~\cite{gem5}. Table~\ref{tab:parameters} lists
the main simulation parameters. We assume that application data is
mapped such that all bitwise operations happen across aligned-rows
within a subarray. We quantitatively evaluate three applications:
1)~a database bitmap index~\cite{oracle,redis,rlite,fastbit},
2)~BitWeaving~\cite{bitweaving}, a mechanism to accelerate
database column scan operations, and 3)~a bitvector-based
implementation of the widely-used \emph{set} data structure.  In
Section~\ref{sec:other-apps}, we discuss four other potential
applications that can benefit from Buddy.

\begin{table}[h]\small
  \centering
  \begin{tabular}{ll}
    \toprule
    \multirow{2}{*}{Processor} & x86, 8-wide, out-of-order, 4~Ghz\\
    & 64 entry  instruction queue\\
    L1 cache & 32~KB D-cache, 32~KB I-cache, LRU policy\\
    L2 cache & 2~MB, LRU policy, 64~B cache line size\\
    Main memory & DDR4-2400, 1-channel, 1-rank, 16 banks\\
    \bottomrule
  \end{tabular}
  \caption{Major simulation parameters}
  \label{tab:parameters}
\end{table}

\subsection{Bitmap Indices}
\label{sec:bitmap-indices}

Bitmap indices~\cite{bmide} are an alternative to traditional
B-tree indices for databases. Compared to B-trees, bitmap indices
1)~consume less space, and 2)~can perform better for many
important queries (e.g., joins, scans). Several major databases
support bitmap indices (e.g., Oracle~\cite{oracle},
Redis~\cite{redis}, Fastbit~\cite{fastbit},
rlite~\cite{rlite}). Several real applications (e.g.,
Spool~\cite{spool}, Belly~\cite{belly},
bitmapist~\cite{bitmapist}, Audience Insights~\cite{ai}) use
bitmap indices for fast analytics.  As bitmap indices heavily rely
on bulk bitwise operations, Buddy can accelerate bitmap indices,
thereby improving overall application performance.

\begin{figure}[h]
  \centering
  \includegraphics{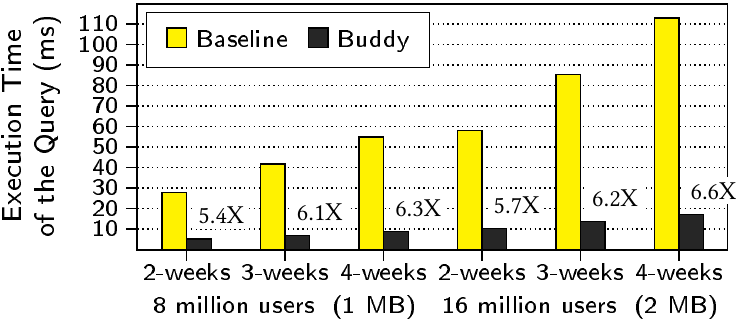}
  \caption{Performance of Buddy for bitmap indices The values on
    top of each bar indicates the factor reduction in execution
    time due to Buddy.}
  \label{fig:rlite}
\end{figure}

To demonstrate this benefit, we use the following workload from a
real application~\cite{ai}. The application uses bitmap indices to
track users' characteristics (e.g., gender) and activities (e.g.,
did the user log in to the website on day 'X'?)  for $m$
users. The applications then uses bitwise operations on these
bitmaps to answer different queries. Our workload runs the
following query: ``How many unique users were active every week
for the past $n$ weeks? and How many male users were active each
of the past $n$ weeks?''  Executing this query requires 6$n$
bitwise \bor, 2$n$-1 bitwise \band, and $n$+1 bitcount
operations. In our mechanism, Buddy accelerates the bitwise \bor
and \band operations in these queries, and the bitcount operations
are performed by the CPU.  Figure~\ref{fig:rlite} shows the
end-to-end query execution time of the baseline and Buddy for the
above experiment for various values of $m$ and $n$.

We draw two conclusions. First, as each query has $O(n)$ bitwise
operations and each bitwise operation takes $O(m)$ time, the
execution time of the query increases with increasing value
$mn$. Second, Buddy significantly reduces the query execution time
compared to the baseline, by 6X on average.

While we demonstrate the benefits of Buddy using one query, as all
bitmap index queries involve several bitwise operations, Buddy
would provide similar performance benefits for any application
using bitmap indices.

\subsection{BitWeaving: Fast Scans using Bitwise Operations}
\label{sec:bitweaving}

Column scan operations are a common part of many database
queries. They are typically performed as part of evaluating a
predicate. For a column with integer values, a predicate is
typically of the form, \texttt{c1 <= val <= c2}. Recent
works~\cite{bitweaving,vectorizing-column-scans} have observed
that existing data representations for storing columnar data are
inefficient for such predicate evaluation especially when the
number of bits used to store each value of the column is less than
the processor width. This is because 1)~the values do not align
well with the processor boundaries, and 2)~the processor typically
does not have comparison instructions at granularities smaller
than the processor word. To address this problem,
BitWeaving~\cite{bitweaving} proposes two different data
representations called BitWeaving-H and BitWeaving-V. We focus our
attention on the faster of the two mechanisms, BitWeaving-V.

BitWeaving-V stores the values of a column such that the first bit
of all the values of the column are stored contiguously, the
second bit of all the values of the column are stored
contiguously, and so on. Using this representation, the predicate
\texttt{c1 <= val <= c2}, can be represented as a series of
bitwise operations starting from the most significant bit all the
way to the least significant bit (we refer the reader to the
BitWeaving paper~\cite{bitweaving} for the detailed algorithm). As
these bitwise operations can be performed in parallel across
multiple values of the column, BitWeaving uses the hardware SIMD
support accelerate these operations. With support for Buddy, these
operations can be performed in parallel across a much larger set
of values, thereby enabling higher performance.

\begin{figure}[h]
  \centering
  \includegraphics{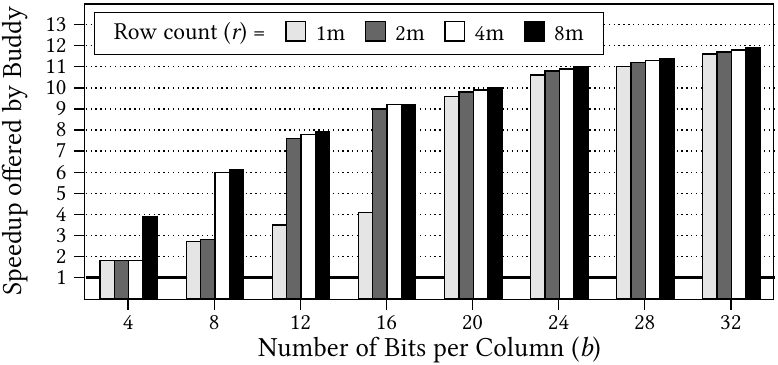}
  \caption{Speedup offered by Buddy for BitWeaving}
  \label{fig:bitweaving}
\end{figure}

We show this benefit by comparing the performance of the baseline
BitWeaving with the performance of BitWeaving accelerated by Buddy
for a commonly-used query\\ \centerline{\small{`\texttt{select count(*)
    from T where c1 <= val <= c2}'}} The query involves a series of
bitwise operations to evaluate the predicate and a bitcount
operation to compute the \texttt{count(*)}. The execution time of
this query depends on 1)~the number of bits used to represent each
value of \texttt{val}, \textit{b}, and 2)~the number of rows in
the table \texttt{T}, \textit{r}. Figure~\ref{fig:bitweaving}
shows the speedup of Buddy over BitWeaving for various values of
\emph{b} and \emph{r}.

We draw three conclusions. First, Buddy improves the performance
of the query by between 1.8X and 11.8X (7.0X on average) compared
to the baseline BitWeaving for various values of \emph{b} and
\emph{r}. Second, the performance improvement of Buddy increases
with increasing value of the number of bits per column \emph{b},
because, as \emph{b} increases, the fraction of time spent in
performing the bitcount operation reduces. As a result, a larger
fraction of the execution can be accelerated using Buddy. Third,
for \emph{b} = 4, 8, 12, and 16, we can observe a large jump in
the speedup of Buddy. These are points where the working set stops
fitting in the on-chip cache. By exploiting the high bank-level
parallelism in DRAM, Buddy can outperform baseline BitWeaving (by
up to 4.1X) \emph{even when} the working set fits in the cache.

\subsection{Bit Vectors vs. Red-Black Trees}
\label{sec:bitset}

Many algorithms heavily use the \emph{set} data structure. While
red-black trees~\cite{red-black-tree} (RB-trees) are commonly used
to implement a set (e.g., C++ Standard Template
Library~\cite{stl}), when the domain of elements is limited, we
can implement a set using a bit vector. Bit vectors offer constant
time \emph{insert} and \emph{lookup} as opposed to RB-trees, which
consume $O(\log n)$ time for both operations. However, with bit
vectors, set operations like \emph{union}, \emph{intersection},
and \emph{difference} have to operate on the entire bit vector,
regardless of whether the elements are actually present in the
set. As a result, for these operations, depending on the number of
elements in the set, bit vectors may outperform or perform worse
than RB-trees. With support for fast bulk bitwise operations, we
show that Buddy significantly shifts the trade-off in favor of bit
vectors.

\begin{figure}[h]
  \centering
  \includegraphics{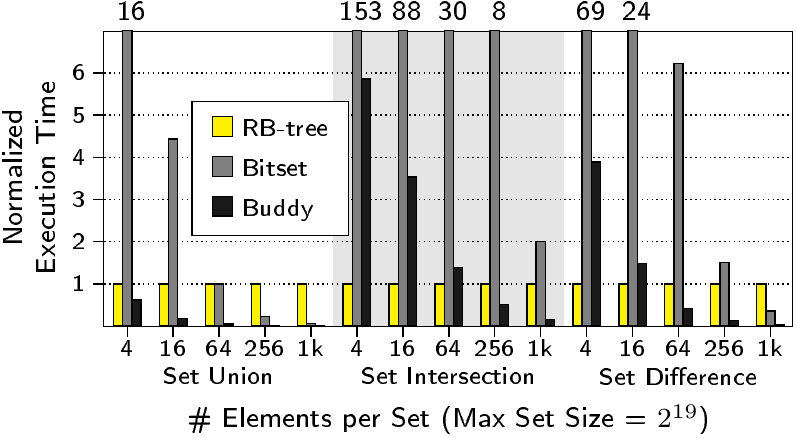}
  \caption{Comparison between RB-Tree, Bitset, and Buddy}
  \label{plot:set-results}
\end{figure}

To demonstrate this, we compare the performance of \emph{union},
\emph{intersection}, and \emph{difference} operations using three
implementations: RB-tree, bit vectors with SIMD optimization
(Bitset), and bit vectors with Buddy. We run a benchmark that
performs each operation on $k$ sets (containing elements between 1
and $2^{19}$) and stores the result in an output set.
Figure~\ref{plot:set-results} shows the execution time for each
implementation normalized to RB-tree for the three operations for
$k = 15$ with varying number of elements in the input sets.

We draw three conclusions. First, by enabling much higher
throughput for bitwise operations, Buddy outperforms the baseline
bitset on all the experiments. Second, as expected, when the
number of elements in each set is very small (16 out of $2^{19}$),
RB-Tree performs better than the bit vector based
implementations. Third, even when each set contains only 64 or
more (out of $2^{19}$) elements, Buddy significantly outperforms
RB-Tree, 3X on average.

In summary, by performing bulk bitwise operations efficiently and
with much higher throughput compared to existing systems, Buddy
makes a bit-vector-based implementation of a set more attractive
than red-black-trees.

\subsection{Other Applications}
\label{sec:other-apps}

\subsubsection{Masked Initialization.}

Certain operations have to clear a specific field in an array of
objects. Such \emph{masked initializations}~\cite{intel-mmx} are
very useful in applications like graphics (e.g., clearing a
specific color in an image). Existing systems read the entire data
structure into the processor/GPU before performing these
operations. Fortunately, we can use bitwise AND/OR operations to
express masked initializations, and consequently, Buddy can easily
accelerate these operations.

\subsubsection{Encryption.} Many encryption algorithms  heavily
use bitwise operations (e.g., XOR)~\cite{xor1,xor2,enc1}. The
Buddy support for fast and efficient bitwise operations can
i)~boost the performance of existing encryption algorithms, and
ii)~enable new encryption algorithms with high throughput and
efficiency.

\subsubsection{DNA Sequence Mapping.}  Most DNA sequence mapping
algorithms~\cite{dna-overview} rely on identifying the locations
where a small DNA sub-string occurs in the reference genome. As
the reference genome is large, prior works have proposed a number
of pre-processing
algorithms~\cite{dna-algo1,dna-algo2,dna-algo3,dna-algo4,shd,dna-our-algo,bitwise-alignment}
have to speedup this operation. Some of these prior
works~\cite{shd,dna-our-algo,bitwise-alignment} heavily using
bitwise operations. Buddy can significantly accelerate these
bitwise operations, thereby reducing the overall time consumed by
the DNA sequence mapping algorithm.

\subsubsection{Approximate Statistics.} Certain large systems employ
probabilistic data structures to improve the efficiency of
maintaining statistics~\cite{summingbird}. Many such structures
(e.g., Bloom filters) rely on bitwise operations to achieve high
efficiency. By improving the throughput of bitwise operations,
Buddy can improve the efficiency of such data structures, and
potentially enable the design of new data structures in this
space.

%% file: related.tex
\section{Related Work}
\label{sec:related}

To our knowledge, this is the first work that proposes a mechanism
to perform the functionally complete set of bitwise operations
completely inside DRAM with high efficiency \emph{and} low
cost. While other works have explored using capacitors to
implement logic gates~\cite{threshold-logic}, we are not aware of
any work that exploits modern DRAM architecture to perform bitwise
operations. Many prior works aim to enable efficient computation
near memory. We now compare Buddy to these prior works.

Some recent patents~\cite{mikamonu,mikamonu2} from Mikamonu
describe a new DRAM organization with 3T-1C cells and additional
logic (e.g., muxes) to perform NAND/NOR operations on the data
inside DRAM. While this architecture can perform bitwise
operations inside DRAM, the 3T-1C cells results in
\emph{significant additional area cost} to the DRAM array, and
hence greatly reduces overall memory density/capacity. In
contrast, Buddy exploits existing DRAM cell structure and
operation to perform bitwise operations efficiently inside
DRAM. As a result, it incurs much lower cost compared to the
Mikamonu architecture.

A recent paper proposes Pinatubo~\cite{pinatubo}, a mechanism to
perform bulk bitwise operations inside PCM.  Similarly, a recent
line of
work~\cite{memristor-1,memristor-2,memristor-3,memristor-4}
proposes mechanisms to perform bitwise operations and other simple
operations (3-bit full adder) completely inside a memristor
array. First, as the underlying memory technology is different,
the mechanisms proposed by these works is completely different
from Buddy. Second, given that DRAM is faster than PCM or
memristor, Buddy can offer much higher throughput compared to
Pinatubo. Having said that, these works demonstrate the importance
of improving the throughput of bulk bitwise operations.

A few recent works~\cite{rowclone,isaac,sramsod} exploit memory
architectures to accelerate specific operations.
RowClone~\cite{rowclone} efficiently performs bulk copy and
initialization inside DRAM. Kang et al.~\cite{sramsod} propose a
mechanism to exploit SRAM to accelerate ``sum of absolute
differences'' computation. ISAAC~\cite{isaac} proposes a mechanism
to accelerate dot-product computations using a memristor
array. While these mechanisms significantly improve the efficiency
of performing the respective operations, none of them can perform
bitwise operations like Buddy.

Prior works (e.g., EXECUBE~\cite{execube}, IRAM~\cite{iram},
DIVA~\cite{diva}) propose designs that \emph{integrate custom
  processing logic} into the DRAM chip to perform bandwidth
intensive operations. The idea is to design processing elements
using the DRAM process, thereby allowing these elements to exploit
the full bandwidth of DRAM.  These approaches have two
drawbacks. First, logic designed using DRAM process is generally
slower than regular processors. Second, the logic added to DRAM
significantly increases the chip cost. In contrast, we propose a
low-cost mechanism that greatly accelerates bitwise operations.

Some recent DRAM architectures~\cite{3d-stacking,hmc,hbm} use
3D-stacking to stack multiple DRAM chips on top of the processor
chip or a separate logic layer. These architectures offer much
higher bandwidth to the logic layer compared to traditional
off-chip interfaces. This enables an opportunity to offload some
computation to the logic layer, thereby improving performance. In
fact, many recent works have proposed mechanisms to improve and
exploit such architectures
(e.g.,~\cite{pim-enabled-insts,pim-graph,top-pim,nda,msa3d}). Even
though they higher bandwidth compared to off-chip memory, such
3D-stacked architectures are still bandwidth
limited~\cite{smla}. However, since Buddy can be integrated easily
with such architectures, it can still offer significant
performance and energy improvements in conjunction with
3D-stacking.

%% file: conclusion.tex
\section{Conclusion}
\label{sec:conclusion}

We introduced Buddy, a new substrate that performs row-wide
bitwise operations within a DRAM chip by exploiting the analog
operation of DRAM. Buddy consists of two components. The first
component uses simultaneous activation of three DRAM rows to
efficiently perform bitwise AND/OR operations. The second
component uses the inverters present in each sense amplifier to
efficiently implement bitwise NOT operations. With these two
components, Buddy can perform \emph{any} bitwise logical operation
efficiently within DRAM. Our evaluations show that Buddy enables
10.9X--25.6X improvement in the throughput of bitwise
operations. This improvement directly translates to significant
performance improvement (3X--7X) in the evaluated data-intensive
applications.  Buddy is generally applicable to any memory
architecture that uses DRAM technology. We believe that the
support for fast and efficient bitwise operations in DRAM can
enable better design of applications to take advantage of them,
which would result in large improvements in performance and
efficiency.

%% file: buddy.bbl
\begin{thebibliography}{10}

\bibitem{intel-skylake}
{6th Generation Intel Core Processor Family Datasheet}.
\newblock
  \url{http://www.intel.com/content/www/us/en/processors/core/desktop-6th-gen-core-family-datasheet-vol-1.html}.

\bibitem{belly}
Belly card engineering.
\newblock \url{https://tech.bellycard.com/}.

\bibitem{bitmapist}
{bitmapist: Powerful realtime analytics with Redis 2.6's bitmaps and Python}.
\newblock \url{https://github.com/Doist/bitmapist}.

\bibitem{stl}
C++ containers libary, std::set.
\newblock \url{http://en.cppreference.com/w/cpp/container/set}.

\bibitem{fastbit}
{FastBit: An Efficient Compressed Bitmap Index Technology}.
\newblock \url{https://sdm.lbl.gov/fastbit/}.

\bibitem{gtx745}
{GeForce GTX 745 Specifications}.
\newblock
  \url{http://www.geforce.com/hardware/desktop-gpus/geforce-gtx-745-oem/specifications}.

\bibitem{hbm}
{High Bandwidth Memory DRAM}.
\newblock \url{http://www.jedec.org/standards-documents/docs/jesd235}.

\bibitem{redis}
Redis - bitmaps.
\newblock \url{http://redis.io/topics/data-types-intro#bitmaps}.

\bibitem{rlite}
{rlite: A Self-contained, Serverless, Zero-configuration, Transactional
  Redis-compatible Database Engine}.
\newblock \url{https://github.com/seppo0010/rlite}.

\bibitem{spool}
Spool.
\newblock \url{http://www.getspool.com/}.

\bibitem{pim-graph}
Junwhan Ahn, Sungpack Hong, Sungjoo Yoo, Onur Mutlu, and Kiyoung Choi.
\newblock A scalable processing-in-memory accelerator for parallel graph
  processing.
\newblock In {\em Proceedings of the 42Nd Annual International Symposium on
  Computer Architecture}, ISCA '15, pages 105--117, New York, NY, USA, 2015.
  ACM.

\bibitem{pim-enabled-insts}
Junwhan Ahn, Sungjoo Yoo, Onur Mutlu, and Kiyoung Choi.
\newblock Pim-enabled instructions: A low-overhead, locality-aware
  processing-in-memory architecture.
\newblock In {\em Proceedings of the 42Nd Annual International Symposium on
  Computer Architecture}, ISCA '15, pages 336--348, New York, NY, USA, 2015.
  ACM.

\bibitem{mikamonu2}
A.~Akerib, O.~AGAM, E.~Ehrman, and M.~Meyassed.
\newblock {Using storage cells to perform computation}, December 2014.
\newblock US Patent 8,908,465.

\bibitem{mikamonu}
Avidan Akerib and Eli Ehrman.
\newblock {In-memory Computational Device, Patent No. 20150146491}, May 2015.

\bibitem{bitwise-alignment}
Gary Benson, Yozen Hernandez, and Joshua Loving.
\newblock {\em {A Bit-Parallel, General Integer-Scoring Sequence Alignment
  Algorithm}}, pages 50--61.
\newblock Springer Berlin Heidelberg, Berlin, Heidelberg, 2013.

\bibitem{gem5}
Nathan Binkert, Bradford Beckmann, Gabriel Black, Steven~K. Reinhardt, Ali
  Saidi, Arkaprava Basu, Joel Hestness, Derek~R. Hower, Tushar Krishna, Somayeh
  Sardashti, Rathijit Sen, Korey Sewell, Muhammad Shoaib, Nilay Vaish, Mark~D.
  Hill, and David~A. Wood.
\newblock {The Gem5 Simulator}.
\newblock {\em SIGARCH Comput. Archit. News}, 39(2):1--7, August 2011.

\bibitem{summingbird}
Oscar Boykin, Sam Ritchie, Ian O'Connell, and Jimmy Lin.
\newblock Summingbird: A framework for integrating batch and online mapreduce
  computations.
\newblock {\em Proc. VLDB Endow.}, 7(13):1441--1451, August 2014.

\bibitem{bmide}
Chee-Yong Chan and Yannis~E. Ioannidis.
\newblock Bitmap index design and evaluation.
\newblock In {\em SIGMOD}, pages 355--366, New York, NY, USA, 1998. ACM.

\bibitem{dsarp}
K.~K.-W. Chang, D.~Lee, Z.~Chisti, A.~R. Alameldeen, C.~Wilkerson, Y.~Kim, and
  O.~Mutlu.
\newblock {Improving DRAM Performance by Parallelizing Refreshes with
  Accesses}.
\newblock In {\em HPCA}, 2014.

\bibitem{linux-dma}
J.~Corbet et~al.
\newblock {\em Linux Device Drivers}, page 445.
\newblock O'Reilly Media, 2005.

\bibitem{ai}
Demiz Denir, Islam AbdelRahman, Liang He, and Yingsheng Gao.
\newblock {Audience Insights query engine: In-memory integer store for social
  analytics}.
\newblock \texttt{https://code.facebook.com/posts/382299771946304/
  audience-insights-query-engine-in-memory-
  integer-store-for-social-analytics-/}.

\bibitem{diva}
Jeff Draper, Jacqueline Chame, Mary Hall, Craig Steele, Tim Barrett, Jeff
  LaCoss, John Granacki, Jaewook Shin, Chun Chen, Chang~Woo Kang, Ihn Kim, and
  Gokhan Daglikoca.
\newblock {The Architecture of the DIVA Processing-in-memory Chip}.
\newblock In {\em Proceedings of the 16th International Conference on
  Supercomputing}, ICS '02, pages 14--25, New York, NY, USA, 2002. ACM.

\bibitem{nda}
A.~Farmahini-Farahani, Jung~Ho Ahn, K.~Morrow, and Nam~Sung Kim.
\newblock Nda: Near-dram acceleration architecture leveraging commodity dram
  devices and standard memory modules.
\newblock In {\em High Performance Computer Architecture (HPCA), 2015 IEEE 21st
  International Symposium on}, pages 283--295, Feb 2015.

\bibitem{red-black-tree}
Leo~J. Guibas and Robert Sedgewick.
\newblock {A Dichromatic Framework for Balanced Trees}.
\newblock In {\em Proceedings of the 19th Annual Symposium on Foundations of
  Computer Science}, SFCS '78, pages 8--21, Washington, DC, USA, 1978. IEEE
  Computer Society.

\bibitem{msa3d}
Qi~Guo, Nikolaos Alachiotis, Berkin Akin, Fazle Sadi, Guanglin Xu, Tze~Meng
  Low, Larry Pileggi, James~C Hoe, and Franz Franchetti.
\newblock {3D-stacked Memory-side Acceleration: Accelerator and System Design}.
\newblock In {\em WoNDP}, 2013.

\bibitem{xor2}
Jong-Wook Han, Choon-Sik Park, Dae-Hyun Ryu, and Eun-Soo Kim.
\newblock {Optical image encryption based on XOR operations}.
\newblock {\em Optical Engineering}, 38(1):47--54, 1999.

\bibitem{charge-cache}
H.~Hassan, G.~Pekhimenko, N.~Vijaykumar, V.~Seshadri, D.~Lee, O.~Ergin, and
  O.~Mutlu.
\newblock {ChargeCache: Reducing DRAM latency by exploiting row access
  locality}.
\newblock In {\em 2016 IEEE International Symposium on High Performance
  Computer Architecture (HPCA)}, pages 581--593, March 2016.

\bibitem{intel-avx}
Intel.
\newblock {Intel Instruction Set Architecture Extensions}.
\newblock \url{https://software.intel.com/en-us/intel-isa-extensions}.

\bibitem{hmc}
J.~Jeddeloh and B.~Keeth.
\newblock {Hybrid Memory Cube: New DRAM architecture increases density and
  performance}.
\newblock In {\em VLSIT}, pages 87--88, June 2012.

\bibitem{ddr3-1600}
JEDEC.
\newblock {DDR3 SDRAM Standard, JESD79-3D}.
\newblock \url{http://www.jedec.org/sites/default/files/docs/JESD79-3D.pdf},
  2009.

\bibitem{2t-1c-1}
H.B. Kang and S.K. Hong.
\newblock One-transistor type dram, January~8 2009.
\newblock US Patent App. 12/000,393.

\bibitem{sramsod}
Mingu Kang, Min-Sun Keel, Naresh~R Shanbhag, Sean Eilert, and Ken Curewitz.
\newblock {An energy-efficient VLSI architecture for pattern recognition via
  deep embedding of computation in SRAM}.
\newblock In {\em 2014 IEEE International Conference on Acoustics, Speech and
  Signal Processing (ICASSP)}, pages 8326--8330. IEEE, 2014.

\bibitem{uksong-kang}
Uksong Kang.
\newblock {Personal communication}, Oct 2016.

\bibitem{dram-cd}
Brent Keeth, R.~Jacob Baker, Brian Johnson, and Feng Lin.
\newblock {\em DRAM Circuit Design: Fundamental and High-Speed Topics}.
\newblock Wiley-IEEE Press, 2nd edition, 2007.

\bibitem{salp}
Yoongu Kim, Vivek Seshadri, Donghyuk Lee, Jamie Liu, and Onur Mutlu.
\newblock {A Case for Exploiting Subarray-level Parallelism (SALP) in DRAM}.
\newblock In {\em Proceedings of the 39th Annual International Symposium on
  Computer Architecture}, ISCA '12, pages 368--379, Washington, DC, USA, 2012.
  IEEE Computer Society.

\bibitem{btt-knuth}
D.~E. Knuth.
\newblock {The Art of Computer Programming. Fascicle 1: Bitwise Tricks \&
  Techniques; Binary Decision Diagrams}, 2009.

\bibitem{execube}
Peter~M. Kogge.
\newblock {EXECUBE: A New Architecture for Scaleable MPPs}.
\newblock In {\em ICPP}, pages 77--84, Washington, DC, USA, 1994. IEEE Computer
  Society.

\bibitem{memristor-2}
S.~Kvatinsky, D.~Belousov, S.~Liman, G.~Satat, N.~Wald, E.~G. Friedman,
  A.~Kolodny, and U.~C. Weiser.
\newblock {MAGIC ---Memristor-Aided Logic}.
\newblock {\em IEEE Transactions on Circuits and Systems II: Express Briefs},
  61(11):895--899, Nov 2014.

\bibitem{memristor-3}
S.~Kvatinsky, A.~Kolodny, U.~C. Weiser, and E.~G. Friedman.
\newblock {Memristor-based IMPLY logic design procedure}.
\newblock In {\em Computer Design (ICCD), 2011 IEEE 29th International
  Conference on}, pages 142--147, Oct 2011.

\bibitem{memristor-4}
S.~Kvatinsky, G.~Satat, N.~Wald, E.~G. Friedman, A.~Kolodny, and U.~C. Weiser.
\newblock {Memristor-Based Material Implication (IMPLY) Logic: Design
  Principles and Methodologies}.
\newblock {\em IEEE Transactions on Very Large Scale Integration (VLSI)
  Systems}, 22(10):2054--2066, Oct 2014.

\bibitem{al-dram}
D.~Lee, Y.~Kim, G.~Pekhimenko, S.~Khan, V.~Seshadri, K.~Chang, and O.~Mutlu.
\newblock {Adaptive-latency DRAM: Optimizing DRAM timing for the common-case}.
\newblock In {\em HPCA}, pages 489--501, Feb 2015.

\bibitem{smla}
Donghyuk Lee, Saugata Ghose, Gennady Pekhimenko, Samira Khan, and Onur Mutlu.
\newblock {Simultaneous Multi-Layer Access: Improving 3D-Stacked Memory
  Bandwidth at Low Cost}.
\newblock {\em ACM Trans. Archit. Code Optim.}, 12(4):63:1--63:29, January
  2016.

\bibitem{tl-dram}
Donghyuk Lee, Yoongu Kim, Vivek Seshadri, Jamie Liu, Lavanya Subramanian, and
  Onur Mutlu.
\newblock {Tiered-latency DRAM: A Low Latency and Low Cost DRAM Architecture}.
\newblock In {\em Proceedings of the 2013 IEEE 19th International Symposium on
  High Performance Computer Architecture (HPCA)}, HPCA '13, pages 615--626,
  Washington, DC, USA, 2013. IEEE Computer Society.

\bibitem{memristor-1}
Yifat Levy, Jehoshua Bruck, Yuval Cassuto, Eby~G. Friedman, Avinoam Kolodny,
  Eitan Yaakobi, and Shahar Kvatinsky.
\newblock {Logic operations in memory using a memristive Akers array}.
\newblock {\em Microelectronics Journal}, 45(11):1429 -- 1437, 2014.

\bibitem{dna-algo1}
Heng Li and Richard Durbin.
\newblock {Fast and accurate long-read alignment with Burrows--Wheeler
  transform}.
\newblock {\em Bioinformatics}, 26(5):589--595, 2010.

\bibitem{pinatubo}
Shuangchen Li, Cong Xu, Qiaosha Zou, Jishen Zhao, Yu~Lu, and Yuan Xie.
\newblock {Pinatubo: A Processing-in-Memory Architecture for Bulk Bitwise
  Operations in Emerging Non-Volatile Memories}.
\newblock In {\em Proceedings of the 53rd Annual Design Automation Conference},
  page 173. ACM, 2016.

\bibitem{bitweaving}
Yinan Li and Jignesh~M. Patel.
\newblock {BitWeaving: Fast Scans for Main Memory Data Processing}.
\newblock In {\em Proceedings of the 2013 ACM SIGMOD International Conference
  on Management of Data}, SIGMOD '13, pages 289--300, New York, NY, USA, 2013.
  ACM.

\bibitem{data-retention}
Jamie Liu, Ben Jaiyen, Yoongu Kim, Chris Wilkerson, and Onur Mutlu.
\newblock {An Experimental Study of Data Retention Behavior in Modern DRAM
  Devices: Implications for Retention Time Profiling Mechanisms}.
\newblock In {\em Proceedings of the 40th Annual International Symposium on
  Computer Architecture}, ISCA '13, pages 60--71, New York, NY, USA, 2013. ACM.

\bibitem{3d-stacking}
Gabriel~H. Loh.
\newblock {3D-Stacked Memory Architectures for Multi-core Processors}.
\newblock In {\em Proceedings of the 35th Annual International Symposium on
  Computer Architecture}, ISCA '08, pages 453--464, Washington, DC, USA, 2008.
  IEEE Computer Society.

\bibitem{enc1}
S.~A. Manavski.
\newblock {CUDA Compatible GPU as an Efficient Hardware Accelerator for AES
  Cryptography}.
\newblock In {\em IEEE International Conference on Signal Processing and
  Communications, 2007. ICSPC 2007}, pages 65--68, Nov 2007.

\bibitem{bmidc}
Elizabeth O'Neil, Patrick O'Neil, and Kesheng Wu.
\newblock Bitmap index design choices and their performance implications.
\newblock In {\em Proceedings of the 11th International Database Engineering
  and Applications Symposium}, IDEAS '07, pages 72--84, Washington, DC, USA,
  2007. IEEE Computer Society.

\bibitem{oracle}
{Oracle}.
\newblock {Using Bitmap Indexes in Data Warehouses}.
\newblock
  \url{https://docs.oracle.com/cd/B28359_01/server.111/b28313/indexes.htm}.

\bibitem{threshold-logic}
H.~Ozdemir, A.~Kepkep, B.~Pamir, Y.~Leblebici, and U.~Cilingiroglu.
\newblock {A capacitive threshold-logic gate}.
\newblock {\em IEEE Journal of Solid-State Circuits}, 31(8):1141--1150, Aug
  1996.

\bibitem{iram}
David Patterson, Thomas Anderson, Neal Cardwell, Richard Fromm, Kimberly
  Keeton, Christoforos Kozyrakis, Randi Thomas, and Katherine Yelick.
\newblock {A Case for Intelligent RAM}.
\newblock {\em IEEE Micro}, 17(2):34--44, March 1997.

\bibitem{intel-mmx}
Alex Peleg and Uri Weiser.
\newblock {MMX Technology Extension to the Intel Architecture}.
\newblock {\em IEEE Micro}, 16(4):42--50, August 1996.

\bibitem{ptmweb}
PTM.
\newblock Predictive technology model.
\newblock http://ptm.asu.edu/.

\bibitem{rambus}
Rambus.
\newblock {DRAM Power Model}.
\newblock \url{https://www.rambus.com/energy/}, 2010.

\bibitem{dna-our-algo}
Kim~R Rasmussen, Jens Stoye, and Eugene~W Myers.
\newblock {Efficient q-gram filters for finding all $\varepsilon$-matches over
  a given length}.
\newblock {\em Journal of Computational Biology}, 13(2):296--308, 2006.

\bibitem{vrt-2}
P.~J. Restle, J.~W. Park, and B.~F. Lloyd.
\newblock {DRAM variable retention time}.
\newblock In {\em International Electron Devices Meeting, 1992. IEDM '92.
  Technical Digest}, pages 807--810, Dec 1992.

\bibitem{dna-algo2}
Stephen~M Rumble, Phil Lacroute, Adrian~V Dalca, Marc Fiume, Arend Sidow, and
  Michael Brudno.
\newblock {SHRiMP: Accurate mapping of short color-space reads}.
\newblock 2009.

\bibitem{dna-overview}
Sophie Schbath, V{\'e}ronique Martin, Matthias Zytnicki, Julien Fayolle,
  Valentin Loux, and Jean-Fran{\c{c}}ois Gibrat.
\newblock {Mapping reads on a genomic sequence: An algorithmic overview and a
  practical comparative analysis}.
\newblock {\em Journal of Computational Biology}, 19(6):796--813, 2012.

\bibitem{dbi}
Vivek Seshadri, Abhishek Bhowmick, Onur Mutlu, Phillip~B. Gibbons, Michael~A.
  Kozuch, and Todd~C. Mowry.
\newblock {The Dirty-block Index}.
\newblock In {\em Proceeding of the 41st Annual International Symposium on
  Computer Architecuture}, ISCA '14, pages 157--168, Piscataway, NJ, USA, 2014.
  IEEE Press.

\bibitem{rowclone}
Vivek Seshadri, Yoongu Kim, Chris Fallin, Donghyuk Lee, Rachata
  Ausavarungnirun, Gennady Pekhimenko, Yixin Luo, Onur Mutlu, Phillip~B.
  Gibbons, Michael~A. Kozuch, and Todd~C. Mowry.
\newblock {RowClone: Fast and Energy-efficient in-DRAM Bulk Data Copy and
  Initialization}.
\newblock In {\em Proceedings of the 46th Annual IEEE/ACM International
  Symposium on Microarchitecture}, MICRO-46, pages 185--197, New York, NY, USA,
  2013. ACM.

\bibitem{isaac}
Ali Shafiee, Anirban Nag, Naveen Muralimanohar, Rajeev Balasubramonian,
  John~Paul Strachan, Miao Hu, R~Stanley Williams, and Vivek Srikumar.
\newblock {ISAAC: A Convolutional Neural Network Accelerator with In-Situ
  Analog Arithmetic in Crossbars}b.
\newblock In {\em Proc. ISCA}, 2016.

\bibitem{nuat}
W.~Shin, J.~Yang, J.~Choi, and L.~S. Kim.
\newblock {NUAT: A non-uniform access time memory controller}.
\newblock In {\em 2014 IEEE 20th International Symposium on High Performance
  Computer Architecture (HPCA)}, pages 464--475, Feb 2014.

\bibitem{xor1}
P.~Tuyls, H.~D.~L. Hollmann, J.~H.~Van Lint, and L.~Tolhuizen.
\newblock {XOR-based Visual Cryptography Schemes}.
\newblock {\em Designs, Codes and Cryptography}, 37(1):169--186.

\bibitem{hacker-delight}
Henry~S. Warren.
\newblock {\em Hacker's Delight}.
\newblock Addison-Wesley Professional, 2nd edition, 2012.

\bibitem{dna-algo3}
David Weese, Anne-Katrin Emde, Tobias Rausch, Andreas D{\"o}ring, and Knut
  Reinert.
\newblock {RazerS—fast read mapping with sensitivity control}.
\newblock {\em Genome research}, 19(9):1646--1654, 2009.

\bibitem{vectorizing-column-scans}
Thomas Willhalm, Ismail Oukid, Ingo Müller, and Franz Faerber.
\newblock {Vectorizing Database Column Scans with Complex Predicates}.
\newblock In Rajesh Bordawekar, Christian~A. Lang, and Bugra Gedik, editors,
  {\em ADMS@VLDB}, pages 1--12, 2013.

\bibitem{bicompression}
Kesheng Wu, Ekow~J. Otoo, and Arie Shoshani.
\newblock {Compressing Bitmap Indexes for Faster Search Operations}.
\newblock In {\em Proceedings of the 14th International Conference on
  Scientific and Statistical Database Management}, SSDBM '02, pages 99--108,
  Washington, DC, USA, 2002. IEEE Computer Society.

\bibitem{shd}
Hongyi Xin, John Greth, John Emmons, Gennady Pekhimenko, Carl Kingsford, Can
  Alkan, and Onur Mutlu.
\newblock {Shifted Hamming Distance: A Fast and Accurate SIMD-Friendly Filter
  to Accelerate Alignment Verification in Read Mapping}.
\newblock {\em Bioinformatics}, 2015.

\bibitem{dna-algo4}
Hongyi Xin, Donghyuk Lee, Farhad Hormozdiari, Samihan Yedkar, Onur Mutlu, and
  Can Alkan.
\newblock {Accelerating read mapping with FastHASH}.
\newblock {\em BMC genomics}, 14(Suppl 1):S13, 2013.

\bibitem{vrt-1}
D.~S. Yaney, C.~Y. Lu, R.~A. Kohler, M.~J. Kelly, and J.~T. Nelson.
\newblock {A meta-stable leakage phenomenon in DRAM charge storage - Variable
  hold time}.
\newblock In {\em 1987 International Electron Devices Meeting}, pages 336--339,
  Dec 1987.

\bibitem{top-pim}
Dongping Zhang, Nuwan Jayasena, Alexander Lyashevsky, Joseph~L. Greathouse,
  Lifan Xu, and Michael Ignatowski.
\newblock Top-pim: Throughput-oriented programmable processing in memory.
\newblock In {\em Proceedings of the 23rd International Symposium on
  High-performance Parallel and Distributed Computing}, HPDC '14, pages 85--98,
  New York, NY, USA, 2014. ACM.

\bibitem{half-dram}
Tao Zhang, Ke~Chen, Cong Xu, Guangyu Sun, Tao Wang, and Yuan Xie.
\newblock {Half-DRAM: A High-bandwidth and Low-power DRAM Architecture from the
  Rethinking of Fine-grained Activation}.
\newblock In {\em Proceeding of the 41st Annual International Symposium on
  Computer Architecuture}, ISCA '14, pages 349--360, Piscataway, NJ, USA, 2014.
  IEEE Press.

\bibitem{zhaoptm}
Wei Zhao and Yu~Cao.
\newblock New generation of predictive technology model for sub-45 nm early
  design exploration.
\newblock {\em IEEE TED}, 2006.

\end{thebibliography}
